\documentclass[a4paper,11pt]{article}
\usepackage{lmodern}

\usepackage[T1]{fontenc}
\usepackage[latin9]{inputenc}
\usepackage{mathrsfs}
\usepackage{amsmath}
\usepackage{graphicx}
\PassOptionsToPackage{normalem}{ulem}
\usepackage{ulem}

\makeatletter
\newcommand{\lyxaddress}[1]{
	\par {\raggedright #1
	\vspace{1.4em}
	\noindent\par}
}

\usepackage{cancel}
\usepackage{mathrsfs}   %\mathscr{L}
\usepackage{slashed}     %\slashed{p}
\usepackage{bbold}  % \mathbb{1} for the identity matrix
\usepackage{url}
\usepackage{graphicx}
\usepackage[colorlinks=true,linkcolor=redLinks,citecolor=greenLinks,urlcolor=redLinks, pdfborder={0 0 1}]{hyperref}
\usepackage{xcolor}
\usepackage{framed}
\usepackage[numbers,sort&compress]{natbib}
\usepackage{amsmath}
\usepackage{ytableau}
\usepackage{microtype}
\usepackage[htt]{hyphenat}
\usepackage{titlesec}
\usepackage{booktabs}

\allowdisplaybreaks

\colorlet{shadecolor}{gray!15}

\definecolor{greenLinks}{rgb}{0, 0.6, 0} 
\definecolor{blueLinks}{rgb}{0, 0, 0.6}
\definecolor{redLinks}{rgb}{0.6, 0, 0}
\definecolor{tempText}{rgb}{0.55, 0.10,0.67}
\definecolor{eprintLinks}{rgb}{0.4, 0.4, 0.4}
\definecolor{journalLinks}{rgb}{0.6, 0, 0}

\titleformat{\section}[block]{\color{black}\Large\bfseries\filcenter}{\thetitle.\;}{0em}{}
\titleformat{\subsection}[block]{\color{black}\large\bfseries\filcenter}{}{0em}{}

\newcommand{\MYhref}[3][redLinks]{\href{#2}{\color{#1}{#3}}}%

\usepackage{multirow}
\let\orig@Hy@EveryPageAnchor\Hy@EveryPageAnchor
\def\Hy@EveryPageAnchor{%
    \begingroup
    \hypersetup{pdfview=Fit}%
    \orig@Hy@EveryPageAnchor
    \endgroup
}

\let\oldFootnote\footnote
\newcommand\nextToken\relax

\renewcommand\footnote[1]{%
    \oldFootnote{#1}\futurelet\nextToken\isFootnote}

\newcommand\isFootnote{%
    \ifx\footnote\nextToken\textsuperscript{,}\fi}

\usepackage[most]{tcolorbox} 

\definecolor{block-gray}{gray}{0.95}
\definecolor{darkRed}{rgb}{0.67, 0, 0}

\newtcolorbox{codeExample}{
    enhanced,
    frame hidden,
    colback=block-gray,
    boxrule=0pt,
    borderline west={2pt}{0pt}{gray!80!black}
}

\newtcolorbox{codeSyntax}{
	collower=black,
	bicolor,
	colback=gray!80!black,
	colupper=white,
	colbacklower=block-gray,
	colframe=gray!80!black,
	boxrule=2pt,
	fontupper=\ttfamily\bfseries,
	sharp corners=all
    }

\newtcolorbox{codeSyntaxRed}{
	collower=black,
	bicolor,
	colback=red!75!black,
	colupper=white,
	colbacklower=red!5!white,
	colframe=red!75!black,
	boxrule=2pt,
	fontupper=\ttfamily\bfseries,
	sharp corners=all
    }

\makeatother

\begin{document}
\title{Using \texttt{SimTeEx} to simplify polynomial expressions with tensors}
\author{Renato M. Fonseca\date{}}
\maketitle

\lyxaddress{\begin{center}
{\Large{}\vspace{-0.5cm}}High Energy Theory Group\\
Departamento de Física Teórica y del Cosmos,\\
Universidad de Granada, E--18071 Granada, Spain\\
~\\
Email: renatofonseca@ugr.es
\par\end{center}}
\begin{abstract}
Computations with tensors are ubiquitous in fundamental physics, and
so is the usage of Einstein's dummy index convention for the contraction
of indices. For instance, $T_{ia}U_{aj}$ is readily recognized as
the same as $T_{ib}U_{bj}$, but a computer does not know that \texttt{T{[}i,a{]}U{[}a,j{]}}
is equal to \texttt{T{[}i,b{]}U{[}b,j{]}}. Furthermore, tensors may
have symmetries which can be used to simply expressions: if $U_{ij}$
is antisymmetric, then $\alpha T_{ia}U_{aj}+\beta T_{ib}U_{jb}=\left(\alpha-\beta\right)T_{ia}U_{aj}$.
The fact that tensors can have elaborate symmetries, together with
the problem of dummy indices, makes it complicated to simplify polynomial
expressions with tensors. In this work I will present an algorithm
for doing so, which was implemented in the Mathematica package \texttt{SimTeEx}
(\textit{\uline{Sim}}\textit{plify }\textit{\uline{Te}}\textit{nsor
}\textit{\uline{Ex}}\textit{pressions}). It can handle any kind
of tensor symmetry.
\end{abstract}

\section{\label{sec:Introduction}Introduction}

In particle physics as well and in general relativity one often has
to deal with expressions involving tensors, such as the Riemann tensor
or the Wilson coefficients of an effective field theory. The indices
of such tensors appear so often contracted that implicit summation
of repeated indices \cite{Einstein:1916vd} is by now second nature
to researchers in these fields.

Index contractions and the potential symmetries under exchange of
indices can make it non-trivial to simplify expressions involving
polynomials of tensors. For example the symmetries of the Riemann
tensor,
\begin{equation}
R_{ijkl}=-R_{jikl}\,,R_{ijkl}=-R_{ijlk}\textrm{ and }R_{ijkl}+R_{iklj}+R_{iljk}=0\,,\label{eq:R-sym}
\end{equation}
imply that \cite{Peeters:2018dyg}
\begin{align}
R_{pqrs}R_{ptru}R_{tvqw}R_{uvsw}-R_{pqrs}R_{pqtu}R_{rvtw}R_{svuw}\nonumber \\
-R_{mnab}R_{npbc}R_{mscd}R_{spda}+\frac{1}{4}R_{mnab}R_{psba}R_{mpcd}R_{nsdc} & =0\,.\label{eq:R-relation}
\end{align}
Likewise, in the Standard Model effective field theory (SMEFT), at
dimension 6, one encounters the operator \cite{Weinberg:1979sa,Wilczek:1979hc,Abbott:1980zj}
\begin{equation}
\mathcal{O}_{ijkl}=\epsilon_{\alpha\beta\gamma}\epsilon_{nm}\epsilon_{pq}\left(Q_{i,\alpha n}^{T}CQ_{j,\beta p}\right)\left(Q_{k,\gamma q}^{T}CL_{l,m}\right)
\end{equation}
where the external indices $\left(ijkl\right)$ label the fields'
flavor. The remaining subscripts are $SU(3)$ and $SU(2)$ indices,
which are unimportant for the present discussion. Crucially, $\mathcal{O}_{ijkl}$
is a tensor with the non-trivial symmetry \cite{Abbott:1980zj}
\begin{align}
\mathcal{O}_{ijkl}+\mathcal{O}_{jikl}-\mathcal{O}_{kijl}-\mathcal{O}_{kjil} & =0\,.\label{eq:Op-QQQL-sym}
\end{align}
It follows that the Wilson coefficient $\kappa_{ijkl}$ which contracts
with $\mathcal{O}_{ijkl}$ in the SMEFT Lagrangian is the most general
tensor obeying the slightly different relation
\begin{align}
	\kappa_{ijkl}+\kappa_{jikl}-\kappa_{jkil}-\kappa_{kjil} & =0 \label{eq:QQQL-sym}
\end{align}
that leads to complicated
polynomial relations involving $\kappa$, such as 
\begin{align}
\kappa_{a b b m} \kappa_{a c d m} \kappa_{d p p n} \kappa_{q q c n} + \kappa_{a b a m} \kappa_{b c d m} \kappa_{d p p n} \kappa_{q q c n}\nonumber \\+ 2 \kappa_{a a b m} \kappa_{b c d m} \kappa_{p d p n} \kappa_{q q c n} -  4 \kappa_{a a b m} \kappa_{b c d m} \kappa_{p p c n} \kappa_{q q d n}  &=0\,.\label{eq:QQQL-relation}
\end{align}

The symmetries of $\kappa$ and the Riemann tensor $R$ and often
called \textit{multi-term} as they involve equations with more than
two terms. One cannot fully account for the peculiar properties of
these tensors by simply tracking some sign (or phase) under exchange
of indices. Another way of looking at these tensors is to say that,
under permutations of their indices, they do not transform as 1-dimension
representations of the relevant permutation group. In fact, it is
well know that $R$ transforms as the\ytableausetup{centertableaux} \ytableausetup{boxsize=0.7em}
\begin{equation}
\ydiagram{2,2}
\end{equation}
irreducible representation of $S_{4}$, which is 2-dimensional. On
the other hand, $\kappa$ transforms as the
\begin{equation}
\left(\ydiagram{3}+\ydiagram{2,1}+\ydiagram{1,1,1}\right)\times\ydiagram{1}\label{eq:k-symmetry}
\end{equation}
reducible representation of $S_{3}\times S_{1}$, which has dimension
$1+2+1=4$. Even more complicated symmetries arise in effective field
theories when one considers operators of higher dimensions, such as
those which violate baryon and/or lepton number --- see for example
\cite{Babu:2001ex,deGouvea:2007qla,Fonseca:2018aav,Gargalionis:2020xvt}.
Having said this, the reader unfamiliar with the permutation group
and its representations shouldn't be overly concerned as the algorithm
described in this paper does not rely on it, not does on need to know
any of this in order to use the main function of the \texttt{SimTeEx}
program (to be introduced later), which puts a tensor expression in
canonical form. The program also contains some auxiliary functions,
described in the appendix \ref{sec:Extra-tools}, which do require
some knowledge of the permutation group. I have tried to make the
discussion there somewhat self-contained, however textbooks on the
matter, such as \cite{Tung-book}, might still come in handy.

The twofold purpose of this work is to (1) describe an algorithm which
can take into account arbitrarily complicated symmetry relations,
producing a canonical form for a polynomial expression with one or
more tensors, and (2) introduce the Mathematica package \texttt{SimTeEx}
which implements these computations. Although the algorithm discussed
in this paper draws no inspiration from previous works, it turns out
that it shares similarities with several previous codes:
\begin{itemize}
\item Dummy indices are dealt with graphs, which are used to represent tensor
monomials in a canonical form, making \texttt{SimTeEx} similar to
\texttt{Redberry} \cite{Bolotin:2013qgr} in this aspect.
\item Multi-term symmetries are seen as a linear algebra problem to be solved
by putting matrices in reduced echelon form, an idea which is akin
to the one used in \texttt{ATENSOR} \cite{Ilyin:1996otf}.
\item Columns of these matrices are reordered with the purpose of making
sure that the number of monomials of an input expression does not
increase in the final result. This is analogous to the \texttt{meld}
algorithm \cite{Price:2022wlt} of \texttt{Cadabra} \cite{Peeters:2018dyg}.
\end{itemize}
To this list of packages, I should add \texttt{xPerm} \cite{Martin-Garcia:2008ysv}
(part of \texttt{xAct} \cite{xAct}), which also is capable of simplifying
tensor expressions.

~

Readers only interested in using the \texttt{SimTeEx} program may
jump directly to section \ref{sec:Using-SimTensor}. For those interested
in the algorithm used by this code,
\begin{itemize}
\item section \ref{sec:Dummy-indices} introduces graphs as a way of handling
dummy indices, and in doing so they can be used to represent each
tensor monomial in a canonical form;
\item section \ref{sec:Polynomials-as-a} briefly discusses the fact that
tensor polynomials are a vector space generated by the above mentioned
graphs;
\item section \ref{sec:Tensors-with-symmetries} addresses the difficulty
introduced by tensors with permutation symmetries, and equates the
task of simplifying expressions with them to putting a matrix in reduced
echelon form.
\end{itemize}
A summary is presented in section \ref{sec:Summary-and-future}, together
with a discussion of possible future developments for the \texttt{SimTeEx}
program. In appendix \ref{sec:Extra-tools} the reader can find a
description of several extra functions which are available in \texttt{SimTeEx},
and finally appendix \ref{sec:Tensor-symmetries-beyond} discusses
why Young symmetrizers may not be sufficient to describe the symmetries
of a tensor.

\section{\label{sec:Dummy-indices}Dummy indices and graphs}

Let us ignore for now the possibilities of tensors having permutation
symmetries. Graphs are a very suggestive way of representing a product
of contracted tensors.\footnote{For the more mathematical inclined readers, I should note that one
can have pairs of vertices connected with multiple edges (see figures
\ref{fig:1} and \ref{fig:2}), and/or edges connecting a vertex to
itself (as in figure \ref{fig:2}). As such, strictly speaking, we
are dealing with multigraphs/pseudographs.} For example, considering (wrongly) for a moment that $R$ is a fully
symmetric rank-4 tensor, the first term in equation (\ref{eq:R-relation})
is quite naturally represented by the 5-loop diagram in figure \ref{fig:1}.
\begin{center}
\begin{figure}[h]
\begin{centering}
\includegraphics[scale=1.35]{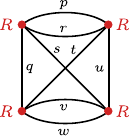}
\par\end{centering}
\caption{\label{fig:1}Potential graph representation of $R_{pqrs}R_{ptru}R_{tvqw}R_{uvsw}$,
assuming that $R$ is completely symmetric. Edge labels are shown
only to make it easier to compare this representation with the original
expression.}
\end{figure}
\par\end{center}

Likewise, under the assumption that $\kappa$ is also fully symmetry
and ignoring the constant prefactor of 2, the first term in equation
(\ref{eq:QQQL-relation}) has the graph representation show in figure
\ref{fig:2}.
\begin{center}
\begin{figure}[h]
\begin{centering}
\includegraphics[scale=1.35]{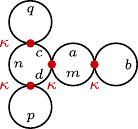}
\par\end{centering}
\caption{\label{fig:2}Potential graph representation of $\kappa_{a b b m} \kappa_{a c d m} \kappa_{d p p n} \kappa_{q q c n}$,
assuming that $\kappa$ is completely symmetric. As in figure \ref{fig:1},
the edges should remain unlabeled (they are only shown here for clarity).}
\end{figure}
\par\end{center}

\begin{figure}[h]
\begin{centering}
\includegraphics[scale=1.35]{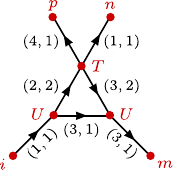}
\par\end{centering}
\caption{\label{fig:3}Directed graph representation of $U_{ijk}U_{klm}T_{njlp}$.
Edge labels $(n_{1},n_{2})$ indicate that the index in slot \#$n_{1}$
of the departing tensor contracts with the index in slot \#$n_{2}$
of the incident tensor. For example the (3,2) implies that the third
index of $T$ contract with the second index of one of the $U$ tensors
(the one with the external $m$ index). It is important to give a
direction to the edges so that one can interpret the two numbers which
label each edge.}
\end{figure}

There are several important observations to be made:
\begin{enumerate}
\item In these two figures, the edges are tagged with the corresponding
dummy indices only to facilitate the comparison with the original
tensor expression. In fact, in order to get a representation of tensor
monomials which is invariant under relabeling of dummy indices, the
graphs edges are to be unlabeled.
\item Not withstanding this last point, I did assume that the participating
tensors ($R$ and $\kappa$) were completely symmetric. When this
is not the case, it is important to differentiate which indices of
the tensors are being contracted, so we do need to label the edges
with the slot positions of the corresponding dummy indices. One can
do so in practice by assigning a direction to the edge, and registering
that information together with the slot positions of the contracting
indices (but not their letters/names).
\item The vertices also need to be labeled with the name of the corresponding
tensor in order to avoid ambiguities. That's because there might be
more than one rank-$n$ tensor in a given monomial.
\item The two example above do not have external indices. When they do exist,
an obvious solution is to treat each of them as a vertex in the diagram,
with degree/valency 1 (i.e. they connect to the rest of the graph
via a single edge).
\end{enumerate}
For instance, $U_{ijk}U_{klm}T_{njlp}$ could be represented by the
directed graph shown in figure \ref{fig:3}, with labeled vertices
and edges.

In this way, the problem of determining if two tensor monomials are
the same is translated to the well known problem of determining if
two graphs are isomorphic.

Let us finish this discussion on the graphical representation of tensor
monomials by briefly observing that obviously, for practical manipulations
such as checking for isomorphisms, one must find some way to represent
the graphs themselves. The program \texttt{SimTeEx} uses what can
be described as a generalized adjacency matrix, which must be significantly
more complicated than a normal one in order to track all the information
mentioned above. Briefly, if we were to remove all the labels and
arrows in figure \ref{fig:3}'s diagram, the resulting graph, with
degree sequence (4,3,3,1,1,1,1), could be represented by the matrix
\begin{equation}
\mathscr{M}=\left(\begin{array}{ccccccc}
0 & 1 & 1 & 1 & 1 & 0 & 0\\
1 & 0 & 1 & 0 & 0 & 1 & 0\\
1 & 1 & 0 & 0 & 0 & 0 & 1\\
1 & 0 & 0 & 0 & 0 & 0 & 0\\
1 & 0 & 0 & 0 & 0 & 0 & 0\\
0 & 1 & 0 & 0 & 0 & 0 & 0\\
0 & 0 & 1 & 0 & 0 & 0 & 0
\end{array}\right)
\end{equation}
whose entry $\left(i,j\right)$ indicates how many connections are
there between vertex $\#i$ and vertex $\#j$. For an undirected graph,
this matrix is symmetric, while for directed ones the entry $\left(i,j\right)$
indicates the number of edges going from vertex $\#i$ to vertex $\#j$,
which might differ from the value of entry $\left(j,i\right)$. To
represent labeled vertices, instead of representing the graph just
with $\mathscr{M}$, one can add a list with the vertex labels (matching
the ordering in the adjacency matrix): $\textrm{graph}=\left\{ \left\{ T,U,U,p,n,i,m\right\} ,\mathscr{M}\right\} $.
As for the edge labels which identify the positions of tensor entries
under contraction, one possibility is the following: if vertex $x$
corresponds to an $r_{x}$-rank tensor, then one can promote entry
$\left(\alpha,\beta\right)$ of $\mathscr{M}$ to an $r_{\alpha}\times r_{\beta}$
dimensional matrix, such that its entry $\left(a,b\right)$ is equal
to 1 if index $\#a$ of the tensor associated to vertex $\alpha$
is contracted with index $\#b$ of the tensor associated to vertex
$\beta$, and 0 otherwise. For example, the monomial $U_{ijk}U_{klm}T_{njlp}$
is fully described by $\left\{ \left\{ T,U,U,p,n,i,m\right\} ,\mathscr{M}^{(\textrm{gen})}\right\} $,
with \arraycolsep=3.0pt
\begin{equation}
\mathscr{M}^{(\textrm{gen})}=\left(\begin{array}{ccccccc}
\underset{}{\left(\begin{array}{cccc}
0 & 0 & 0 & 0\\
0 & 0 & 0 & 0\\
0 & 0 & 0 & 0\\
0 & 0 & 0 & 0
\end{array}\right)} & \left(\begin{array}{ccc}
0 & 0 & 0\\
0 & 1 & 0\\
0 & 0 & 0\\
0 & 0 & 0
\end{array}\right) & \left(\begin{array}{ccc}
0 & 0 & 0\\
0 & 0 & 0\\
0 & 1 & 0\\
0 & 0 & 0
\end{array}\right) & \left(\begin{array}{c}
0\\
0\\
0\\
1
\end{array}\right) & \left(\begin{array}{c}
1\\
0\\
0\\
0
\end{array}\right) & \left(\begin{array}{c}
0\\
0\\
0\\
0
\end{array}\right) & \left(\begin{array}{c}
0\\
0\\
0\\
0
\end{array}\right)\\
\underset{}{\left(\begin{array}{cccc}
0 & 0 & 0 & 0\\
0 & 1 & 0 & 0\\
0 & 0 & 0 & 0
\end{array}\right)} & \left(\begin{array}{ccc}
0 & 0 & 0\\
0 & 0 & 0\\
0 & 0 & 0
\end{array}\right) & \left(\begin{array}{ccc}
0 & 0 & 0\\
0 & 0 & 0\\
1 & 0 & 0
\end{array}\right) & \left(\begin{array}{c}
0\\
0\\
0
\end{array}\right) & \left(\begin{array}{c}
0\\
0\\
0
\end{array}\right) & \left(\begin{array}{c}
1\\
0\\
0
\end{array}\right) & \left(\begin{array}{c}
0\\
0\\
0
\end{array}\right)\\
\underset{}{\left(\begin{array}{cccc}
0 & 0 & 0 & 0\\
0 & 0 & 1 & 0\\
0 & 0 & 0 & 0
\end{array}\right)} & \left(\begin{array}{ccc}
0 & 0 & 1\\
0 & 0 & 0\\
0 & 0 & 0
\end{array}\right) & \left(\begin{array}{ccc}
0 & 0 & 0\\
0 & 0 & 0\\
0 & 0 & 0
\end{array}\right) & \left(\begin{array}{c}
0\\
0\\
0
\end{array}\right) & \left(\begin{array}{c}
0\\
0\\
0
\end{array}\right) & \left(\begin{array}{c}
0\\
0\\
0
\end{array}\right) & \left(\begin{array}{c}
0\\
0\\
1
\end{array}\right)\\
\underset{}{\left(\begin{array}{cccc}
0 & 0 & 0 & 1\end{array}\right)} & \left(\begin{array}{ccc}
0 & 0 & 0\end{array}\right) & \left(\begin{array}{ccc}
0 & 0 & 0\end{array}\right) & \left(\begin{array}{c}
0\end{array}\right) & \left(\begin{array}{c}
0\end{array}\right) & \left(\begin{array}{c}
0\end{array}\right) & \left(\begin{array}{c}
0\end{array}\right)\\
\underset{}{\left(\begin{array}{cccc}
1 & 0 & 0 & 0\end{array}\right)} & \left(\begin{array}{ccc}
0 & 0 & 0\end{array}\right) & \left(\begin{array}{ccc}
0 & 0 & 0\end{array}\right) & \left(\begin{array}{c}
0\end{array}\right) & \left(\begin{array}{c}
0\end{array}\right) & \left(\begin{array}{c}
0\end{array}\right) & \left(\begin{array}{c}
0\end{array}\right)\\
\underset{}{\left(\begin{array}{cccc}
0 & 0 & 0 & 0\end{array}\right)} & \left(\begin{array}{ccc}
1 & 0 & 0\end{array}\right) & \left(\begin{array}{ccc}
0 & 0 & 0\end{array}\right) & \left(\begin{array}{c}
0\end{array}\right) & \left(\begin{array}{c}
0\end{array}\right) & \left(\begin{array}{c}
0\end{array}\right) & \left(\begin{array}{c}
0\end{array}\right)\\
\underset{}{\left(\begin{array}{cccc}
0 & 0 & 0 & 0\end{array}\right)} & \left(\begin{array}{ccc}
0 & 0 & 0\end{array}\right) & \left(\begin{array}{ccc}
0 & 0 & 1\end{array}\right) & \left(\begin{array}{c}
0\end{array}\right) & \left(\begin{array}{c}
0\end{array}\right) & \left(\begin{array}{c}
0\end{array}\right) & \left(\begin{array}{c}
0\end{array}\right)
\end{array}\right)\,.\label{eq:M-example}
\end{equation}
\arraycolsep=5.0ptTaking a look at the block $\left(\alpha,\beta\right)=\left(1,3\right)$
of this large matrix,\arraycolsep=3.0pt
\begin{equation}
\mathscr{M}_{1,3}^{(\textrm{gen})}=\left(\begin{array}{ccc}
0 & 0 & 0\\
0 & 0 & 0\\
0 & 1 & 0\\
0 & 0 & 0
\end{array}\right)\,,
\end{equation}
\arraycolsep=5.0ptone can see how vertex $\alpha=1$ (a $T$ tensor)
contracts with vertex $\beta=3$ (a $U$ tensor): the `1' in the $\left(a,b\right)=\left(3,2\right)$
entry implies that the 3rd index of the $T$ tensor contracts to
the 2nd index of the $U$ tensor.

Since there are only two vertices with the same tensor name, the 2nd
and the 3rd in the list above, the equivalence of $U_{ijk}U_{klm}T_{njlp}$
with some other $U_{...}U_{...}T_{....}$ monomial can be checked
with $2!=2$ row and column permutations of the generalized adjacency
matrix. Rather than do this for every pair of graphs $g=\left\{ \textrm{\ensuremath{\left\langle \textrm{tensor list }\mathscr{T}\right\rangle }},\textrm{\ensuremath{\left\langle \textrm{generalized adjacency matrix }\mathscr{M}^{(\textrm{gen})}\right\rangle }}\right\} $
and $g^{\prime}=\left\{ \textrm{\ensuremath{\left\langle \textrm{tensor list }\mathscr{T}\right\rangle }},\textrm{\ensuremath{\left\langle \textrm{generalized adjacency matrix }\mathscr{M}^{\prime(\textrm{gen})}\right\rangle }}\right\} $
to be compared, it is more practical to put each graph $g$ in a canonical
form $g_{\textrm{canonical}}$ by performing all row/column permutations
leading to equivalent graphs and picking out from this list a representative.\footnote{For example, one can do this by first sorting the list of tensors
$\mathscr{T}$ (and reordering accordingly the rows and columns of
$\mathscr{M}^{(\textrm{gen})}$ while doing so). Then, from all the
equivalent $g_{\pi_{i}}=\left\{ \textrm{\ensuremath{\left\langle \textrm{sorted tensor list }\mathscr{T}\right\rangle }},\mathscr{M}_{\pi_{i}}^{(\textrm{gen})}\right\} $
one can take $g_{\textrm{canonical}}$ to be the one associated to
the smallest $\mathscr{M}_{\pi_{i}}^{(\textrm{gen})}$, under some
sorting criteria.} Two monomials are then equal if and only if their representative
is the same.

\section{Polynomials as a vector space\label{sec:Polynomials-as-a}}

Quite naturally, one can see a polynomial $\boldsymbol{P}$ involving
tensors as being a linear combination of the graphs $g_{i}$ described
above, with numerical or symbolic coefficients $c_{i}$. The distinguishing
feature of these $c_{i}$ is simply that they do not involve tensors
(i.e. objects with indices), for example the $7x$ in $7xU_{ijk}U_{klm}T_{njlp}$.
Thus, a tensor polynomial $\boldsymbol{P}$ can be seen as a vector
\begin{equation}
\boldsymbol{P}=\sum_{i}c_{i}g_{i}
\end{equation}
in a vector-space spanned by several $g_{i}$, and the discussion
above provides a fail-proof method of checking if any of two elements
of this generating set are equivalent. Indeed, if some $g_{1}$ is
found to be equivalent to some $g_{2}$, we can drop one of these
vectors from the generating set, and add the corresponding coefficients:
\begin{equation}
c_{1}g_{1}+c_{2}g_{2}+\cdots\rightarrow\left(c_{1}+c_{2}\right)g_{1}+\cdots\,.
\end{equation}
As an example, consider the tensor polynomial \texttt{2 m{[}a,q,q,b{]}
T{[}a,b{]} + A m{[}c1,m,m,c2{]} T{[}c1,c2{]}} which one can represent
as $2g_{1}+Ag_{2}$. Given the simplicity of this example, and the
absence of symmetries, it is clear already by eye that the two $g$'s
are equivalent: 
\begin{equation}
g_{1}=g_{2}=\left\{ \left\{ m,T\right\} ,\left(\begin{array}{cc}
\underset{}{\left(\begin{array}{cccc}
0 & 0 & 0 & 0\\
0 & 0 & 1 & 0\\
0 & 1 & 0 & 0\\
0 & 0 & 0 & 0
\end{array}\right)} & \left(\begin{array}{cc}
1 & 0\\
0 & 0\\
0 & 0\\
0 & 1
\end{array}\right)\\
\underset{}{\left(\begin{array}{cccc}
1 & 0 & 0 & 0\\
0 & 0 & 0 & 1
\end{array}\right)} & \left(\begin{array}{cc}
0 & 0\\
0 & 0
\end{array}\right)
\end{array}\right)\right\} \,.
\end{equation}
This means that the original expression can be simplified to a single
term: perhaps \texttt{(2+A) m{[}a,q,q,b{]} T{[}a,b{]}} or \texttt{(2+A)
m{[}c1,m,m,c2{]} T{[}c1,c2{]}}. These two possibilities are quite
natural, as they reuse the index labels provided in the input, however
one should keep in mind that the choice of dummy indices in the final
result is arbitrary. Indeed, one could discard altogether the ones
in the input expression, insisting on putting the dummy indices in
some canonical form --- such as $i_{1},i_{2},\cdots$ --- yielding
expressions of the type \texttt{(2+A) m{[}i1,i2,i2,i3{]} T{[}i1,i3{]}}.

\section{Tensors with symmetries\label{sec:Tensors-with-symmetries}}

The discussion so far concerns exclusively the problem introduced
by dummy indices and how to address it by using graphs to represent
tensor monomials. Things becomes more complicated when there are symmetries
under exchange of indices. 

Before diving into these complications, one ought to distinguish the
`simple' symmetries from the `hard' ones. The simple ones, often called
\textit{monoterm symmetries}, are those which can be accounted for
by tracking just a complex phase $\sigma$ (which often is just a
\textit{$\pm$} sign):
\begin{equation}
T_{\pi\left(i_{1}i_{2}\cdots i_{n}\right)}=\sigma T_{i_{1}i_{2}\cdots i_{n}}\,.\label{eq:monoterm-sym}
\end{equation}
Here $\pi$ stands for some permutation of the indices, and $\sigma$
must be an $m$th root of unity ($\sigma^{m}=1$) with $m$ being
the order of the permutation $\pi$, i.e. $\pi^{m}=e$ (the identity).
Here are four examples:
\begin{align}
T_{ab} & =-T_{ba}\,,\\
T_{abcd} & =T_{badc}\,,\\
T_{abcd} & =T_{bacd}\textrm{ and }T_{abcd}=-T_{abdc}\,,\\
T_{abc} & =\omega T_{bca}\textrm{ with }\omega\equiv\exp\left(\frac{2\pi i}{3}\right)\,.
\end{align}
In each of these cases, $T$ is no longer a general tensor, as it
obeys some symmetry under which permuting the indices gives back the
original tensor, up to some phase factor. It is very significant that
the group describing these permutations is abelian in all four cases:
$Z_{2}$, $Z_{2}$, $Z_{2}\times Z_{2}$ and $Z_{3}$, respectively.
That's because the irreducible representations of abelian groups are
1-dimensional, so their action on the indices of some tensor can always
be reduced to studying a phase (as in equation (\ref{eq:monoterm-sym})).

Incorporating tensors with monoterm symmetries in the graph formalism
is straightforward: two monomials are proportional to each other if
they are represented by equivalent graphs, where equivalence is established
not only by permuting equal vertices but also by permuting the edges
associated to the monoterm symmetries of each tensor (i.e. the vertices
of the graphs). Note that it is necessary to track a phase $\sigma$
every time these edge permutations are performed.

Let us then turn our attention to the `hard' symmetries --- the so-called
\textit{multiterm symmetries} --- which are of the form
\begin{equation}
T_{\pi_{1}\left(i_{1}i_{2}\cdots i_{n}\right)}+T_{\pi_{2}\left(i_{1}i_{2}\cdots i_{n}\right)}+\cdots T_{\pi_{p}\left(i_{1}i_{2}\cdots i_{n}\right)}=0\textrm{ with }p>2\,.
\end{equation}
For concreteness, consider a tensor $T$ with the symmetry 
\begin{equation}
T_{abc}+T_{bca}+T_{cab}=0\label{eq:19}
\end{equation}
and, putting aside the problem of dummy indices, the very basic expression
\begin{equation}
xT_{abc}+yT_{bca}+zT_{cab}\label{eq:20}
\end{equation}
which we wish to simplify. One may establish some ordering of least
to most desired form of a tensor, for example $T_{abc}<T_{bca}<T_{cab}$,
in which case we would prioritize eliminating all instances of $T_{abc}$:
\begin{equation}
xT_{abc}+yT_{bca}+zT_{cab}\rightarrow\left(y-x\right)T_{bca}+\left(z-x\right)T_{cab}\,.\label{eq:xyz}
\end{equation}
From this very basic example, we are already in a position to make
some key remarks:
\begin{enumerate}
\item Clearly, the simplified expression depends on the (arbitrary) order
of preference among the various permutations of $T$.
\item By having any fixed order of preference among the various permutations
of $T$, we may be forced to have a result with more monomials than
we started with. That is what happens in expression (\ref{eq:xyz})
for $x=1$ and $y=z=0$. In other words, in the presence of multiterm
symmetries, putting an expression in a \textit{canonical form} may
not necessarily mean the same as \textit{simplifying} it (see also
\cite{Price:2022wlt}).
\item In the expression $U_{abc}\left(xT_{abc}+yT_{bca}+zT_{cab}\right)$,
where a second tensor $U$ with no symmetries was introduced, all
indices are summed over and as such they can be freely relabeled.
Therefore it is meaningless to even consider an order between $T_{abc}$,
$T_{bca}$ and $T_{cab}$. Nonetheless, one can still order the monomials
$U_{abc}T_{abc}$, $U_{abc}T_{bca}$ and $U_{abc}T_{bca}$, by using
a graph representation (as detailed in section \ref{sec:Dummy-indices}).
\end{enumerate}
With these insights, one may reduce the task of simplifying expressions
with multiterm symmetries to a linear algebra problem. The input is
some tensor polynomial $\boldsymbol{P}=\sum_{i=1}^{k}c_{i}g_{i}$
where each of the $k$ terms is represented by a coefficient $c_{i}$
and a graph $g_{i}$, which we can assume from now on to be in a canonical
form. Multiterm symmetries are relations of the form
\begin{equation}
0=\sum_{i=1}^{k}n_{i}^{(a)}g_{i}
\end{equation}
where the index $a$ accounts for the possible existence of several
relations. Some of the $c_{i}$ coefficients of the original expression
$\boldsymbol{P}$ can be converted to zero by adding to $\boldsymbol{P}$
multiples $\omega_{a}$ of $\sum_{i=1}^{k}n_{i}^{(a)}g_{i}$:
\begin{equation}
\boldsymbol{P}=\sum_{i=1}^{k}c_{i}g_{i}\rightarrow\boldsymbol{P}^{\prime}=\boldsymbol{P}+\sum_{a}\omega_{a}\sum_{i=1}^{k}n_{i}^{(a)}g_{i}\equiv\sum_{i=1}^{k}c_{i}^{\prime}g_{i}\,.
\end{equation}
Many readers will probably consider $\boldsymbol{P}^{\prime}$ to
be in the simplest form when there are as many zero $c_{i}^{\prime}$
coefficient as possible. A rather straightforward and efficient way
of nullifying some $c_{i}^{\prime}$ coefficients is to put the matrix
\begin{equation}
\overset{\begin{array}{ccccc}
\;\; & g_{1} & \;\;g_{2} & \;\cdots & \;g_{k}\end{array}}{\overbrace{\left(\begin{array}{@{}c|cccc@{}}
1 & c_{1} & c_{2} & \cdots & c_{k}\\
\cmidrule[0.4pt]{1-5}0 & n_{1}^{(1)} & n_{2}^{(1)} & \cdots & n_{k}^{(1)}\\
0 & n_{1}^{(2)} & n_{2}^{(2)} & \cdots & n_{k}^{(2)}\\
\vdots & \vdots & \vdots & \ddots
\end{array}\right)}}\label{eq:matrix-to-reduce}
\end{equation}
in reduced row echelon form (RREF) and then take $c_{1}^{\prime}$,
$c_{2}^{\prime}$, ...,$c_{k}^{\prime}$ from columns 2 to $k+1$
of the first line.\footnote{The first column in (\ref{eq:matrix-to-reduce}) is introduced only
to make sure that the first line remains at the top.} Note that the number of null coefficients will depend on the ordering
of the columns of this matrix (i.e. the ordering of the $g_{i}$).
In the case of expression (\ref{eq:20}) and a $T$ tensor with the
symmetry (\ref{eq:19}) we would get from this procedures the coefficients
$\left(c_{1}^{\prime},c_{2}^{\prime},c_{3}^{\prime}\right)=\left(0,y-x,z-x\right)$
for each monomial, assuming an ordering $T_{abc}<T_{bca}<T_{cab}$:
\begin{equation}
\overset{\;\;\;\phantom{-}\;\;T_{abc}\;T_{bca}\;T_{cab}}{\overbrace{\left(\begin{array}{@{}c|ccc@{}}
1 & x\; & y\; & z\\
\cmidrule[0.4pt]{1-4}0 & 1 & 1 & 1
\end{array}\right)}}\overset{\textrm{RREF}}{\longrightarrow}\overset{\phantom{-}\;\;T_{abc}\;\;\;\;\;\;T_{bca}\;\;\;\;\;\;\;\;T_{cab}\;\;}{\overbrace{\left(\begin{array}{@{}c|ccc@{}}
1 & 0\; & y-x\; & z-x\\
\cmidrule[0.4pt]{1-4}0 & \cdots & \cdots & \cdots
\end{array}\right)}}\,.\label{eq:26}
\end{equation}
More symmetries --- of the same tensor or perhaps others --- can
be taken into account by simply adding extra lines to this matrix.
Note that the order of the graphs (i.e. the matrix columns) is important,
and to maximize the number of null coefficients one would have to
test all possible column orderings. Since the number of relevant terms
can be quite large, in general it would not be feasible to test all
column ordering. In our very simple example, using the same ordering
as in (\ref{eq:26}), for $x=0$ and $y=z=1$ we would get no change
in the final result, i.e. $\left(c_{1}^{\prime},c_{2}^{\prime},c_{3}^{\prime}\right)=\left(0,1,1\right)$.
However, swapping the positions of $T_{abc}$ and $T_{bca}$, we would
get a single term:
\begin{equation}
\overset{\;\;\;\phantom{-}\;\;T_{bca}\;T_{abc}\;T_{cab}}{\overbrace{\left(\begin{array}{@{}c|ccc@{}}
1 & 1\; & 0\; & 1\\
\cmidrule[0.4pt]{1-4}0 & 1 & 1 & 1
\end{array}\right)}}\overset{\textrm{RREF}}{\longrightarrow}\overset{\phantom{-}\;\;T_{bca}\;\;\;\;\;T_{abc}\;\;\;\;\;\;\;\;T_{cab}\;\;}{\overbrace{\left(\begin{array}{@{}c|ccc@{}}
1 & 0\; & -1\; & 0\\
\cmidrule[0.4pt]{1-4}0 & \cdots & \cdots & \cdots
\end{array}\right)}}\,.\label{eq:27}
\end{equation}

Setting aside the ambitious goal of always having a minimum number
of monomials, as already mentioned above, there is the related issue
of potentially having more terms in the final result than in the input
expression. However, it is rather simple to avoid this outcome. All
that is needed is for the columns associated to terms which do exist
in the input expression (i.e. those for which $c_{i}\neq0$) to appear
last. In the example $\left(c_{1},c_{2},c_{3}\right)=\left(1,0,0\right)$,
which leads to $\left(c_{1}^{\prime},c_{2}^{\prime},c_{3}^{\prime}\right)=\left(0,-1,-1\right)$,
\begin{equation}
\overset{\;\;\;\phantom{-}\;\;T_{abc}\;T_{bca}\;T_{cab}}{\overbrace{\left(\begin{array}{@{}c|ccc@{}}
1 & 1\; & 0\; & 0\\
\cmidrule[0.4pt]{1-4}0 & 1 & 1 & 1
\end{array}\right)}}\overset{\textrm{RREF}}{\longrightarrow}\overset{\phantom{-}\;\;\;\;\;T_{abc}\;\;\;\;\;T_{bca}\;\;\;\;\;\;T_{cab}\;\;}{\overbrace{\left(\begin{array}{@{}c|ccc@{}}
1 & 0\; & -1\; & -1\\
\cmidrule[0.4pt]{1-4}0 & \cdots & \cdots & \cdots
\end{array}\right)}}\,,\label{eq:27-1}
\end{equation}
 one would swap the first and third columns; from $\left(c_{1},c_{2},c_{3}\right)=\left(0,0,1\right)$
we would get an unchanged result $\left(c_{1}^{\prime},c_{2}^{\prime},c_{3}^{\prime}\right)=\left(0,0,1\right)$:
\begin{equation}
\overset{\;\;\;\phantom{-}\;\;T_{cab}\;T_{bca}\;T_{abc}}{\overbrace{\left(\begin{array}{@{}c|ccc@{}}
1 & 0\; & 0\; & 1\\
\cmidrule[0.4pt]{1-4}0 & 1 & 1 & 1
\end{array}\right)}}\overset{\textrm{RREF}}{\longrightarrow}\overset{\;\phantom{-}\;\;T_{cab}\;\;\;\;\;T_{bca}\;\;\;\;T_{abc}}{\overbrace{\left(\begin{array}{@{}c|ccc@{}}
1 & 0\; & 0\; & 1\\
\cmidrule[0.4pt]{1-4}0 & \cdots & \cdots & \cdots
\end{array}\right)}}\,.\label{eq:27-1-1}
\end{equation}
In more elaborate cases, the $c_{i}^{\prime}$ would not necessarily
be the same as the input $c_{i}$. Crucially, it is impossible for
the number of non-zero coefficients to grow. The end effect of this
reshuffling of columns is similar to the one achieved by the \texttt{meld}
algorithm of \texttt{Cadabra 2} \cite{Price:2022wlt}.

\section{\label{sec:Using-SimTensor}Using \texttt{SimTeEx}}

\subsection{The main function}

Readers which are mainly interested in using the \texttt{SimTeEx}
program may do so by first downloading it from
\begin{center}
\texttt{\href{https://renatofonseca.net/simteex}{renatofonseca.net/simteex}}
\par\end{center}

\noindent then installing and loading the program in Mathematica,\footnote{In order to run the extra functions mentioned in appendix \ref{sec:Extra-tools}
it is also necessary to have \texttt{GroupMath} \cite{Fonseca:2020vke}
installed in the system.}
\noindent \begin{flushleft}
\texttt{}<\hspace{0mm}<\texttt{SimTeEx$\grave{\,\textrm{\,}}$}
\par\end{flushleft}

\noindent and finally calling the function\begin{codeSyntax}
CanonicalForm[<tensor polynomial>,<symmetries>]
\tcblower
Simplifies the given polynomial expression. The second argument (which is optional) should be a list of null expressions encoding the symmetries of the various tensors.
\end{codeSyntax}

For example:\texttt{}

\begin{codeExample}

\includegraphics[scale=0.62]{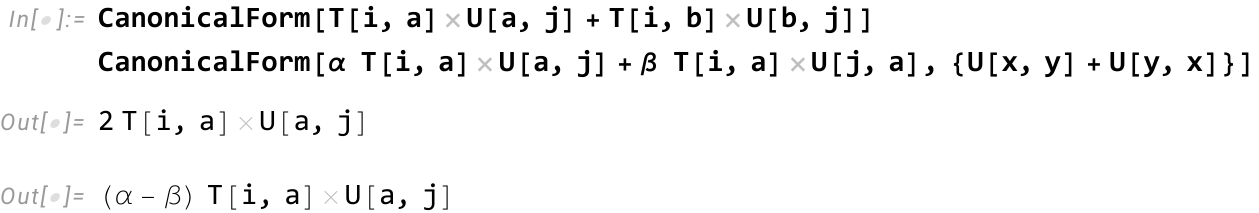}

\end{codeExample}

The last argument in the second example tells the program that the
$U$ tensor has the symmetry $U_{xy}+U_{yx}=0$, i.e. it is anti-symmetric.
The user is free to choose the names of tensors and indices, both
in the expression to simplify as well as in the symmetry conditions.
Everything with a head and square brackets, \texttt{head{[}...{]}},
is assumed to be a tensor, and the variables inside the brackets are
taken to be indices. As such, there is no need to declare the list
of tensors and indices to be used.

Equation (\ref{eq:R-relation}) involving the Riemann tensor can be
checked with the following code:\texttt{}

\begin{codeExample}

\includegraphics[scale=0.62]{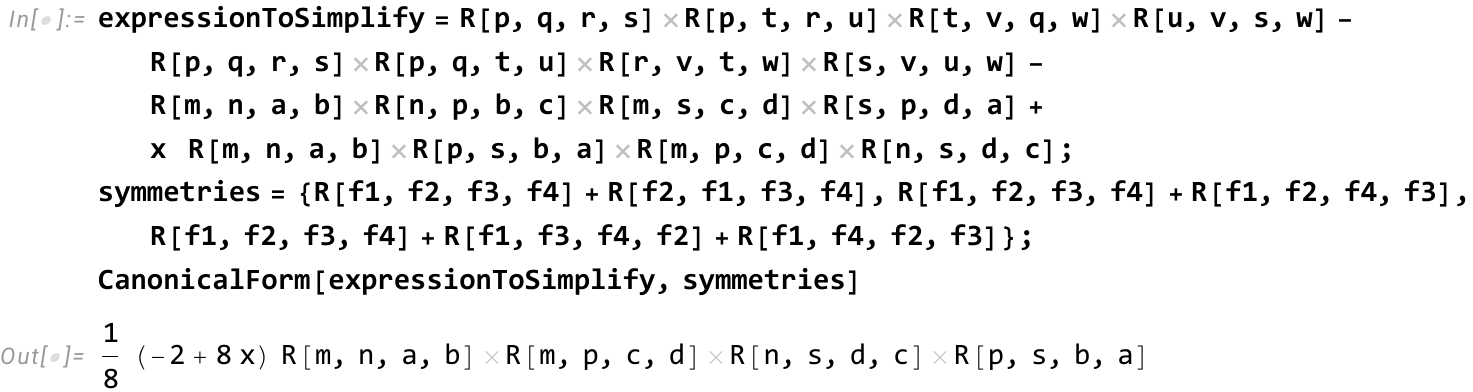}

\end{codeExample}

The factor $1/4$ in (\ref{eq:R-relation}) was intentionally replaced
with a generic $x$ to make clear, from the output, that the expression
is null only for $x=1/4$.

The user does not need to know what is the representation of the permutation
group under which the tensors transform. In fact the user is free
to provide a list of symmetries which makes little sense, such as

\begin{codeExample}

\includegraphics[scale=0.62]{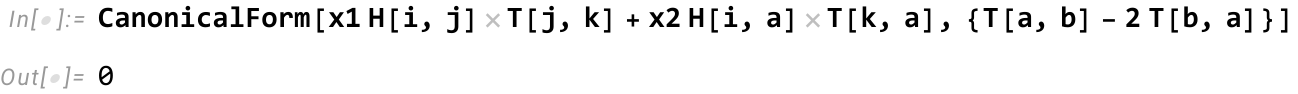}

\end{codeExample}

The zero is explained by the fact that a tensor with the symmetry
$T_{ab}-2T_{ba}=0$ is necessarily null ($T_{ab}=0$). The procedure
described in section \ref{sec:Tensors-with-symmetries} is very flexible,
handling well these situations without any need for special code.

\texttt{Cadabra} \cite{Peeters:2018dyg} --- the only other code
supporting multi-term symmetries which I am aware of --- requires
that the user indicate a Young symmetrizer for each tensor. Asking
only for a set of symmetry equations (like those in equation (\ref{eq:R-sym}))
may offer some advantages. Firstly, while this information is readily
available for well known tensors, in general the user must figure
out the Young tableau associated to a particular set of symmetry relations.
On this topic, I will note that the user is free to provide as input
to \texttt{SimTeEx} any equivalent set of equations (see also the
function \texttt{SameEquationsQ} in appendix \ref{sec:Extra-tools}).
For example, in the case above involving the Riemann tensor, one could
rewrite it as follows:

\begin{codeExample}

\includegraphics[scale=0.62]{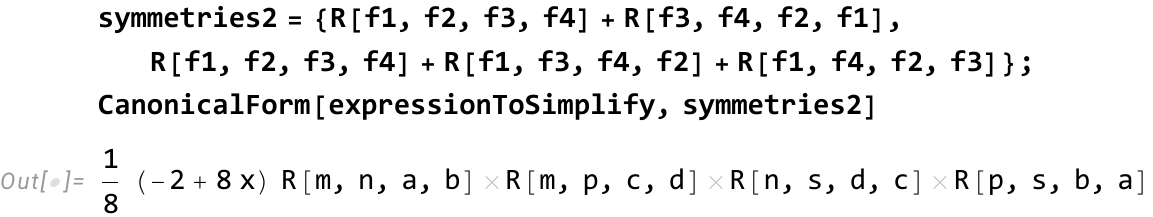}

\end{codeExample}

Secondly, the symmetries of some tensors, such as $\kappa_{ijkl}$
mentioned in the introduction of this work, involve a reducible representation
of the relevant permutation group and therefore they are described
by the sum of several Young symmetrizers. Finally, one can have tensors
with symmetries that cannot be described with Young symmetrizers at
all (readers interested in this topic are referred to appendix \ref{sec:Tensor-symmetries-beyond}
for details).

One should be aware that the function \texttt{CanonicalForm} when
applied to two equivalent expressions, $\textrm{expr}_{1}$ and $\textrm{expr}_{2}$,
may yield different outputs,
\begin{equation}
\mathtt{CanonicalForm}\left(\textrm{expr}_{1}\right)\neq\mathtt{CanonicalForm}\left(\textrm{expr}_{2}\right),\label{eq:canonical-vs-normal-1}
\end{equation}
 although it is always be true that
\begin{equation}
\mathtt{CanonicalForm}\left(\textrm{expr}_{1}-\textrm{expr}_{2}\right)=0\,.\label{eq:canonical-vs-normal-2}
\end{equation}
The reason for the inequality of outputs (\ref{eq:canonical-vs-normal-1})
is twofold:
\begin{enumerate}
\item To ensure that the number of terms never increases, the program performs
the sorting operation which is mentioned at the end of section \ref{sec:Tensors-with-symmetries}.
Since this operation depends on the terms which do appear in the input
one might have $\mathtt{CanonicalForm}\left(\textrm{expr}_{1}\right)\neq\mathtt{CanonicalForm}\left(\textrm{expr}_{2}\right)$
even if the two expressions are the equivalent.
\item An even simpler reason is that \texttt{CanonicalForm} reuses the labels
for dummy indices given in the input, in order not to introduce new
ones which are unfamiliar to the user. Therefore in the trivial example
where $\textrm{expr}_{1}=T_{i}T_{i}$ and $\textrm{expr}_{2}=T_{j}T_{j}$
the function does not change the inputs at all, i.e. $\mathtt{CanonicalForm}\left(\textrm{expr}_{1}\right)=\textrm{expr}_{1}=T_{i}T_{i}$
which is clearly different from $\mathtt{CanonicalForm}\left(\textrm{expr}_{2}\right)=\textrm{expr}_{2}=T_{j}T_{j}$.
On the other hand, $\mathtt{CanonicalForm}\left(T_{i}T_{i}-T_{j}T_{j}\right)=0$.
\end{enumerate}
For some authors \cite{Algorithms-for-computer-algebra,Price:2022wlt}
a function with these properties, (\ref{eq:canonical-vs-normal-1})
and (\ref{eq:canonical-vs-normal-2}), calculates the \textit{normal
form} rather than the \textit{canonical form} of an expression. The
user can change the default behavior of \texttt{CanonicalForm} described
above by setting the global flag \texttt{\$TrueCanonicalForm} to \texttt{True}
(the default value is \texttt{False}):

\begin{codeExample}

\includegraphics[scale=0.62]{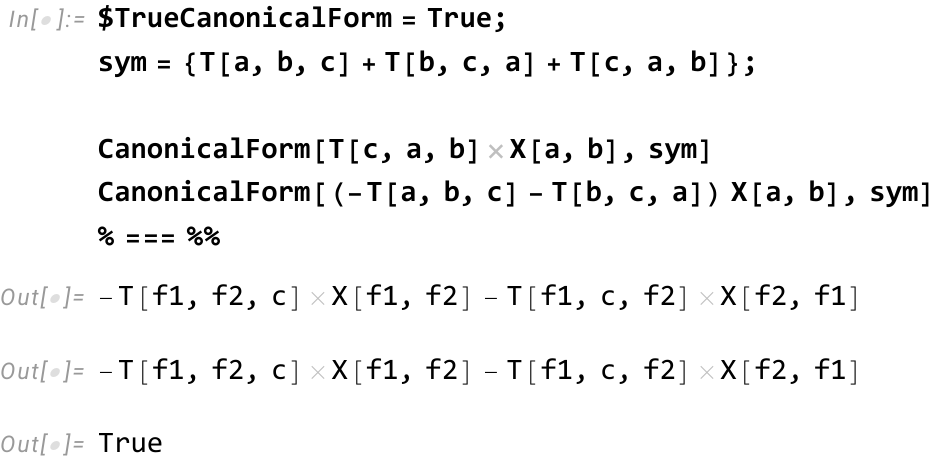}

\end{codeExample}

With \texttt{\$TrueCanonicalForm=True} the dummy indices are picked
from a list in another global flag, \texttt{\$CanonicalListOfIndices},
which the user can change at will (the default value is \texttt{\{f1,f2,f3,f4,} \texttt{f5,f6,...\}}).

\subsection{Expressions with anti-commuting tensors}

In high energy physics one often needs to handle fermion fields which
behave as Grassmann numbers. Being non-trivial representations of
the Lorentz group, they have at the very least a spinor index; very
often they have others, such as gauge and flavor indices. For this
reason, it is useful to consider polynomials where some tensors anti-commute.
\texttt{SimTeEx} is prepared to handle such cases: on one hand, all
fermionic tensors must be declared in the global variable \$\texttt{CanonicalFormFermions};
on the other hand, since the order of the factors is now important,
the built in command \texttt{NonCommutativeMultiply} (\texttt{{*}{*}})
must be used in the expression to simplify.

Consider for example a mass term
\begin{equation}
m_{ij}\epsilon_{\alpha\beta}\psi_{i}^{\alpha}\psi_{j}^{\beta}
\end{equation}
 for Weyl spinors $\psi$, where the $\alpha,\beta=1,2$ are spinor
indices which contract with a Levi-Civita tensor $\epsilon$ to form
a Lorentz invariant, while $i,j$ are flavor indices. It is well known
that the mass matrix $m$ is symmetric (or, more rigorously, that
only its symmetric part contributes to the above expression); this
is only true because the spinors $\psi$ anticommute. One can check
with \texttt{SimTeEx} that $\left(m_{ij}-m_{ji}\right)\epsilon_{\alpha\beta}\psi_{\alpha,i}\psi_{\beta,j}=0$
as follows:

\begin{codeExample}

\includegraphics[scale=0.62]{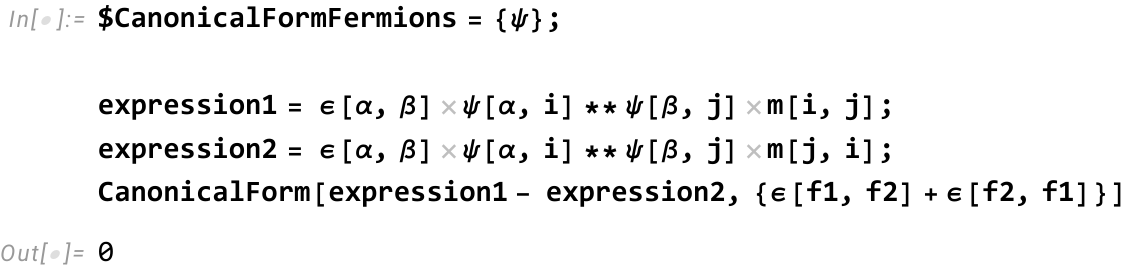}

\end{codeExample}

As a more complicated example, consider the dimension 6 operator obtained
from squaring the one above, i.e.
\begin{equation}
\mathcal{O}_{ijkl}\equiv\left(\psi_{i}^{T}\epsilon\psi_{j}\right)\left(\psi_{k}^{T}\epsilon\psi_{l}\right)=\epsilon_{\alpha\beta}\epsilon_{\gamma\delta}\psi_{i}^{\alpha}\psi_{j}^{\beta}\psi_{k}^{\gamma}\psi_{l}^{\delta}\,.
\end{equation}
Calling $\epsilon_{\alpha\beta\gamma\delta}^{\textrm{sq}}$ to $\epsilon_{\alpha\beta}\epsilon_{\gamma\delta}$,
it is important to note that on top of the obvious symmetries $\epsilon_{\alpha\beta\gamma\delta}^{\textrm{sq}}=-\epsilon_{\beta\alpha\gamma\delta}^{\textrm{sq}}=\epsilon_{\gamma\delta\alpha\beta}^{\textrm{sq}}$,
this tensor also obeys the equation $\epsilon_{\alpha\beta\gamma\delta}^{\textrm{sq}}+\epsilon_{\alpha\delta\beta\gamma}^{\textrm{sq}}+\epsilon_{\alpha\gamma\delta\beta}^{\textrm{sq}}=0$.\footnote{For numerical tensors (such as $\epsilon_{\alpha\beta\gamma\delta}^{\textrm{sq}}$)
one can find systematically all its symmetries by explicitly comparing
the tensor with all its permuted forms; see the function \texttt{SymmetriesOfNumericalTensor}
in subsection (\ref{subsec:SymmetriesOfNumericalTensor}).} Using these properties we can derive that 
\begin{equation}
\mathcal{O}_{ijkl}=\mathcal{O}_{jikl}=\mathcal{O}_{klij}\textrm{ and }\mathcal{O}_{ijkl}+\mathcal{O}_{iklj}+\mathcal{O}_{iljk}=0\,.
\end{equation}

\begin{codeExample}

\includegraphics[scale=0.62]{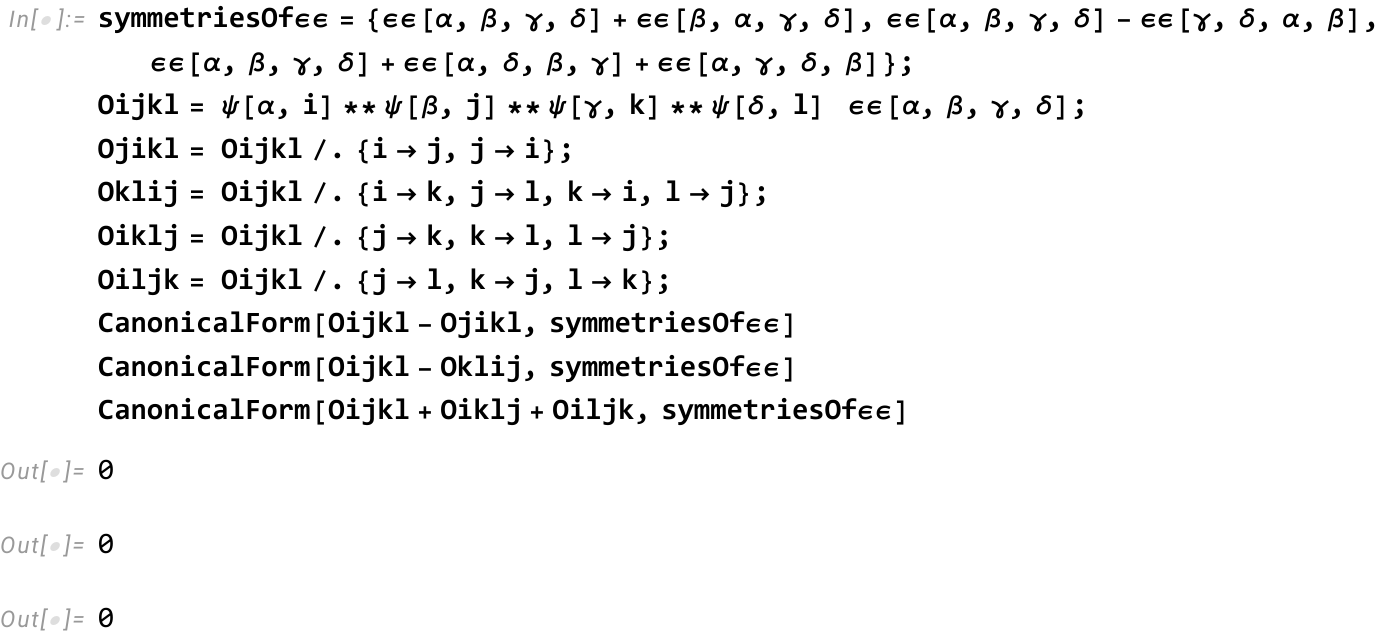}

\end{codeExample}

\subsection{Alternative format for tensor symmetries}

Providing the tensor symmetries as equations is the most general input
format accepted by \texttt{SimTeEx}. However, if so desired, it is
possible to tell the program that some set of indices of a tensor
are fully symmetric or antisymmetric with the alternative format \texttt{\{<tensor
head>,<list of indices>, <1 (for sym) or -1 (for antisym)>\}}:

\begin{codeExample}

\includegraphics[scale=0.62]{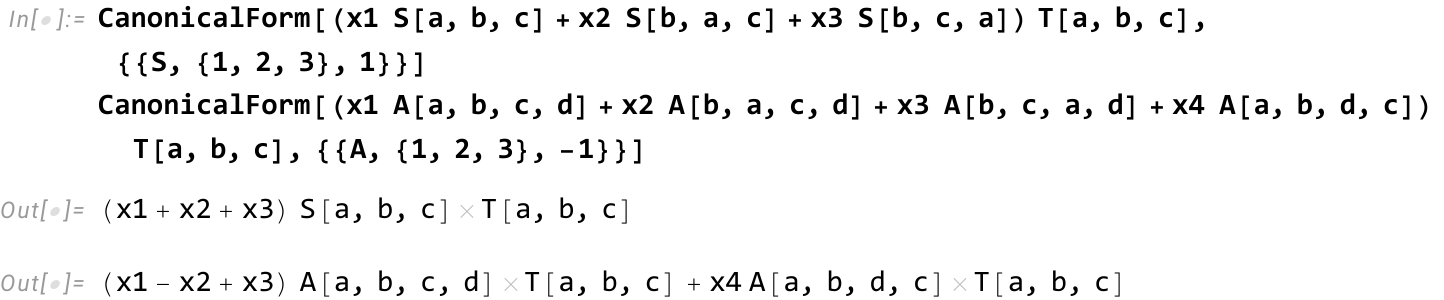}

\end{codeExample}

In the case of a tensor which is not necessarily fully symmetric or
antisymmetric but nonetheless has monoterm symmetries, of the form
in equation (\ref{eq:monoterm-sym}), this can be indicated with the
format \texttt{\{<tensor head>,Cycles{[}...{]}, <phase sigma>\}}:

\begin{codeExample}

\includegraphics[scale=0.62]{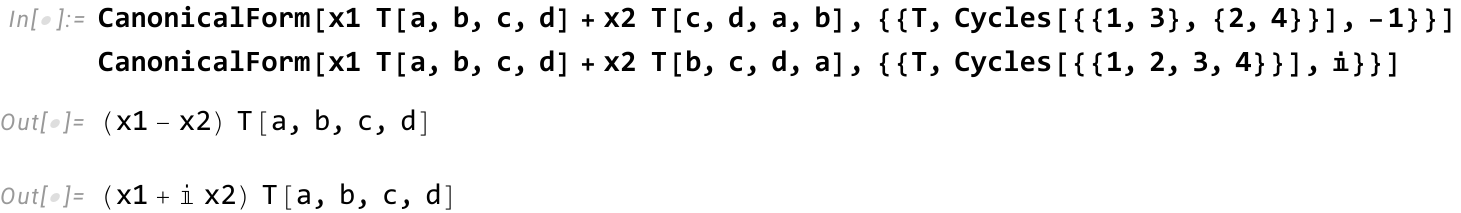}

\end{codeExample}

Internally the code converts these alternative input formats into
a list of symmetry equations.

\subsection{Manually listing the tensor names}

For the user's convenience, the program automatically identifies all
tensors in the given expression by looking for square brackets. This
may sometimes cause problems: for example one cannot use a coefficient
named \texttt{x{[}1{]}} (it would be recognized as a tensor, with
a non-symbolic index, leading to an error). On the other hand, some
quantities with no visible square brackets are internally represented
with them; that is what happens to $\sqrt{2}$, which stands for \texttt{Sqrt{[}2{]}}.
While the particular case of square roots was explicitly addressed
in the code, there might be other objects which create difficulties
to the user. In order to mitigate this issue, the user can manually
provide a list of tensor heads as shown in the following example:

\begin{codeExample}

\includegraphics[scale=0.62]{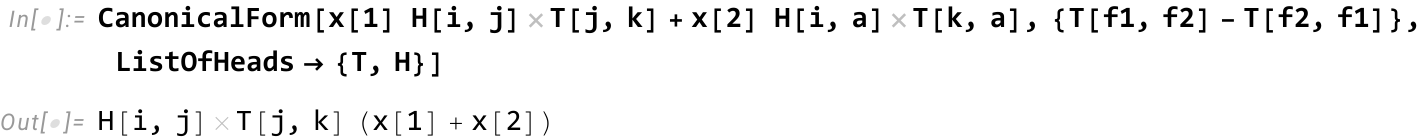}

\end{codeExample}

\section{\label{sec:Summary-and-future}Summary}

On often encounters polynomial expressions with tensors --- some
of which have symmetries --- in several areas of research, such as
in general relativity and in particle physics. It is therefore useful
to have a tool that reliably and automatically simplifies these expressions.
In this work I have presented an algorithm to do so for arbitrarily
complicated tensor symmetries, including the so-called multiterm ones.
It is implemented in the \texttt{CanonicalForm} function of the \texttt{SimTeEx}
Mathematica package, which was designed to be as simple as possible
to use. The current version of the code also contains five extra functions,
described in appendix \ref{sec:Extra-tools},  to analyze, compare
and transform the symmetries of tensors.

The computational performance of the algorithm and the code described
in this work is an important aspect to take into account. Since the
equivalence of two tensor monomials can be equated to finding out
if two graphs are isomorphic, the problem addressed in this paper
is potentially NP-hard. As such, one should not expect to be able
to simplify tensor expressions with too many indices. Nevertheless,
and while speed was not a major design consideration, some simple
modifications to algorithm presented in this work were already implemented
in \texttt{SimTeEx} with the goal of improving the computational time.

\section*{Acknowledgments}

This work was originally developed as a means to expedite calculations
for another research project \cite{Fonseca:workInProgress}, and I
would like to thank José Santiago for testing over and over what eventually
became the \texttt{SimTeEx} code. I am equally grateful to Ricardo
Cepedello and Javi F. Martin for reading draft versions of this text,
and in Ricardo's case also for testing the code. Additionally, I would
like to acknowledge interesting discussions I've had with John Gargalionis
and Anders Eller Thomsen on simplifying tensor expressions, plus thank Zhe Ren and Chang-Yuan Yao for pointing out a mistake in an earlier version of this manuscript.

I acknowledge the financial support from the \textit{Consejería de
Universidad, Investigación e Innovación}, the Spanish government and
the European Union -- NextGenerationEU through grant number AST22\_6.5;
from MCIN/AEI (10.13039/501100011033) through grants number PID2019-106087GB-C22
and PID2022-139466NB-C21; and from the Junta de Andalucía through
grant number P21\_00199 (FEDER).

\appendix

\clearpage

\section{\label{sec:Extra-tools}Extra tools}

Besides the main function --- \texttt{CanonicalForm} --- the \texttt{SimTeEx}
package contains extra code to analyze and process tensors with symmetries.
To ensure that it works properly, the \texttt{GroupMath} \cite{Fonseca:2020vke}
package needs to be installed in the user's computer, in an appropriate
folder, such that it can be loaded automatically by \texttt{SimTeEx}
with the command <\hspace{0mm}<\texttt{GroupMath$\grave{\,\textrm{\,}}$}.

The usage of these extra tools may require some knowledge of the permutation
group $S_{n}$ and its representations. What is mentioned in \cite{Fonseca:2019yya,Fonseca:2020vke}
is sufficient, but in any case, as they become necessary, I lay out
below the most salient group theory aspects to have in mind.

For starters, the reader should recall that the permutation group
of $n$ objects ($S_{n}$) contains $n!$ permutations $\pi$. Each
of them can be expressed in the cycle notation: for example $\pi=\left(132\right)$
is the permutation which replaces object 1 with object 3, object 3
with object 2 and object 2 with object 1: $\left\{ 1,2,3\right\} \rightarrow\left\{ 3,1,2\right\} $.
An algebra is formed when we consider linear combinations $\sum_{\pi\in S_{n}}c_{\pi}\pi$,
as we can multiply ($\star$) two elements of this space using the
group multiplication ($\cdot$), namely $\left(\sum_{\pi\in S_{n}}a_{\pi}\pi\right)\star\left(\sum_{\pi\in S_{n}}b_{\pi}\pi\right)=\sum_{\pi,\pi^{\prime}\in S_{n}}a_{\pi}b_{\pi^{\prime}}\pi\cdot\pi^{\prime}$.

Irreducible representations of $S_{n}$ can be labeled with partitions
of $n$ (e.g. $\left\{ 2,1,1\right\} $ is a partition of $n=4$ since
$2+1+1=4$) which in turn are often depicted graphically as Young
diagrams. In the case of $n=3$, there are three irreducible representations:
$\boldsymbol{1}$(the trivial one), $\boldsymbol{1^{\prime}}$ (the
alternating one) and $\boldsymbol{2}$. They are associated to the
partitions $\{3\}$, $\{1,1,1\}$ and $\{2,1\}$, i.e.
\begin{equation}
\ydiagram{3}\;,\;\;\ydiagram{1,1,1}\;\;\textrm{and}\;\;\ydiagram{2,1}\,.
\end{equation}

\subsection{Young symmetrizers}

\noindent \begin{codeSyntax}
YoungSymmetrizeTensor[<tensor>,<Young tableaux>]
\tcblower
Returns the given tensor projected with the Young symmetrizer associated to the second argument.
\end{codeSyntax}

With this function one can symmetrize a tensor according to some tableaux
$\lambda$. The symmetrizer associated to a Young tableaux can be
defined as follows. Consider the sets of elements of the permutation
group $S_{n}$ which leave the rows and columns of $\lambda$ invariant:
\begin{align}
H_{\lambda} & =\left\{ \pi\in S_{n}:\pi\textrm{ does not change the rows of }\lambda\right\} \,,\\
V_{\lambda} & =\left\{ \pi\in S_{n}:\pi\textrm{ does not change the columns of }\lambda\right\} \,.
\end{align}
Then the Young symmetrizer associated to $\lambda$ is is taken to
be
\begin{equation}
y_{\lambda}\equiv a_{\lambda}s_{\lambda}\textrm{ with }s_{\lambda}=\sum_{h\in H_{\lambda}}h\textrm{ and }a_{\lambda}=\sum_{v\in V_{\lambda}}\textrm{sign}\left(v\right)v\,.
\end{equation}
For an input tensor $T$, the function \texttt{YoungSymmetrizeTensor}
returns $y_{\lambda}T$, reusing the index labels given in the input.
Each tableaux should be provided as lists; for example\ytableausetup{boxsize=1.3em} \begin{align}\begin{ytableau}1 & 3 \\ 2\end{ytableau} &=\mathtt{\{\{1,3\},\{2\}\}}\,,\quad\begin{ytableau}1 & 2\end{ytableau} =\mathtt{\{\{1,2\}\}}\,.\end{align}

\ytableausetup{boxsize=0.7em} 

\begin{codeExample}

\includegraphics[scale=0.62]{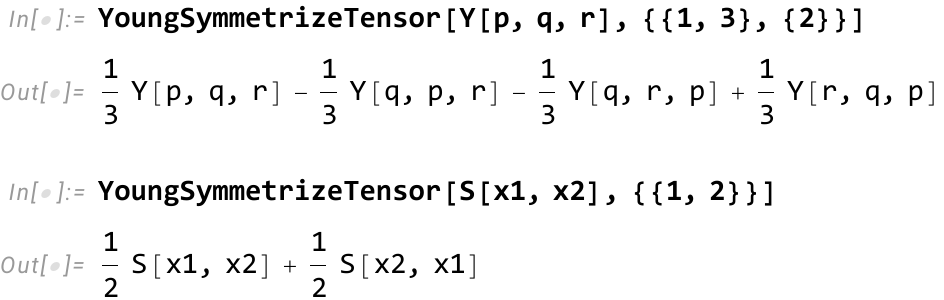}

\end{codeExample}

\subsection{$S_{n}$ irreps in a tensor}

\noindent \begin{codeSyntax}
SnIrrepsInTensor[<null conditions encoding the tensor symmetries>]
\tcblower
Returns the non-null components of a tensor with the given symmetry. The output is a list of the form \{\{partition1, multiplicity1\}, ...\}.
\end{codeSyntax}

Consider a general rank-$n$ tensor, with no symmetries, where all
indices are of the same nature. Assuming that each index can take
$m$ values, one can split the $m^{n}$ components of $T$ according
to how they transform under $S_{n}$ permutations $T_{i_{1}\cdots i_{n}}\rightarrow T_{i_{\pi\left(1\right)}\cdots i_{\pi\left(n\right)}}$.
For an $S_{n}$ irreducible representation labeled by a partition
$\lambda$ of $n$ there are precisely $d\left(\lambda\right)$ parts
of $T$ transforming as $\lambda$, where $d\left(\lambda\right)$
is the dimension of the irreducible representation. For example a
3-index tensor can be split in 4 parts, each transforming according
to the following irreps:
\begin{equation}
\ydiagram{3}+\ydiagram{2,1}+\ydiagram{2,1}+\ydiagram{1,1,1}
\end{equation}
One can use Young symmetrizers to project out each of them. Out of
a total of $m^{n}$, the number of components associated to each piece
can be computed from $m$ and the shape of each diagram $\lambda$,
using a well known formula which is not important for the present
discussion; the interested reader can find more details in \cite{Fonseca:2019yya}
(see also the \texttt{HookContentFormula} function in \cite{Fonseca:2020vke}).

Importantly, in a tensor with symmetries (see appendix \ref{sec:Tensor-symmetries-beyond})
some of these parts are constrained. They could be zero, or perhaps
have relations among themselves; in either case, they are not all
independent. For instance, in a rank-3 fully symmetry tensor there
is only the $\ydiagram{3}$ piece.

Based on a tensor's symmetry equations, the function \texttt{SnIrrepsInTensor}
computes the non-zero components. Here are two examples:

\begin{codeExample}

\includegraphics[scale=0.62]{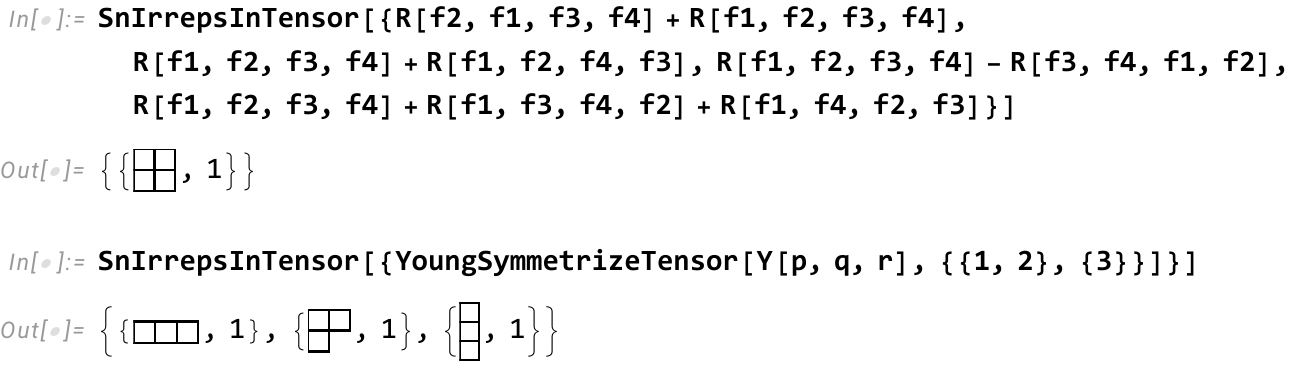}

\end{codeExample}Note that the second case corresponds, for $\lambda=\left\{ \left\{ 1,2\right\} ,\left\{ 3\right\} \right\} $,
to $y_{\lambda}Y$ set to zero (not $y_{\lambda}Y=Y$).

The program will consider that all tensor indices are of the same
nature, always decomposing a rank-$n$ tensor in $S_{n}$ irreps.
Sometimes that might not the best approach: take for example a tensor
$P$ with the symmetry $P_{abcd}=P_{badc}$. Even though the first
two indices never swap with the last two, and therefore one could
consider just the small permutation group $S_{2}\times S_{2}$, currently
the function \texttt{SnIrrepsInTensor} ignores this and decomposes
the tensor in $S_{4}$ irreps:

\begin{codeExample}

\includegraphics[scale=0.62]{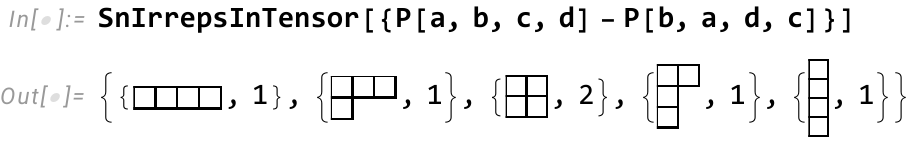}

\end{codeExample}

\subsection{\label{subsec:Condensing-various-tensor}Condensing various tensor
symmetry equations into a single projector}

\begin{codeSyntax}
SingleProjector[<null conditions with the tensor symmetries>]
\tcblower
Returns the unique hermitian projector P such that the condition P(tensor)=tensor is equivalent to the set of null equations given as input.
\end{codeSyntax}

It might sometimes be convenient to reduce a set of symmetry equations
into a single one. The \texttt{SingleProjector} function does that,
by returning a single hermitian operator $P$, which contains all
the input symmetries.\footnote{Note that the input, as always, must be a list $\left\{ \textrm{expr1},\textrm{expr2},...\right\} $
of null expressions, i.e. $\textrm{expr1}=\textrm{expr2}=\cdots=0$,
while the output is a projector $P$ such that $P(\textrm{tensor})=\textrm{tensor}$.
So $\left(P-e\right)(\textrm{tensor})$ is a single null expression
equivalent to the original list.} Requiring that $P$ is a projector ($P^{2}=P$) and hermitian ($P^{\dagger}=P$)
makes it unique.\footnote{The adjoint can be understood as follows. For two tensors $A$ and
$B$ of the same rank, one can define the inner product $\left\langle A,B\right\rangle =A_{i_{1}i_{2}\cdots i_{n}}^{*}B_{i_{1}i_{2}\cdots i_{n}}$.
Then, for some member $\mathcal{X}=\sum_{\pi\in S_{n}}c_{\pi}\pi$
of the $S_{n}$ algebra, where the $c_{\pi}$ are complex numbers,
\[
\left\langle A,\mathcal{X}B\right\rangle \equiv\left\langle \mathcal{X}^{\dagger}A,B\right\rangle =\sum_{\pi\in S_{n}}c_{\pi}A_{i_{1}i_{2}\cdots i_{n}}^{*}B_{\pi\left(i_{1}i_{2}\cdots i_{n}\right)}
\]
 which is the same as $\sum_{\pi\in S_{n}}c_{\pi}A_{\pi^{-1}\left(i_{1}i_{2}\cdots i_{n}\right)}^{*}B_{i_{1}i_{2}\cdots i_{n}}$
or simply $\sum_{\pi\in S_{n}}\left[c_{\pi^{-1}}^{*}A_{\pi\left(i_{1}i_{2}\cdots i_{n}\right)}\right]^{*}B_{i_{1}i_{2}\cdots i_{n}}$.
As such
\[
\mathcal{X}^{\dagger}=\sum_{\pi\in S_{n}}c_{\pi^{-1}}^{*}\pi\,.
\]
For rank-2 tensors, it is well know that if $A$ is (anti)symmetric
then only the (anti)symmetric part of $B$ contributes to the contraction
$A_{ij}B_{ij}$. We are thus entitled to think of $A$ and $B$ as
having the exact same symmetry. However, for higher rank indices this
is only true if $\mathcal{X}=\mathcal{X}^{\dagger}$.}

\begin{codeExample}

\includegraphics[scale=0.62]{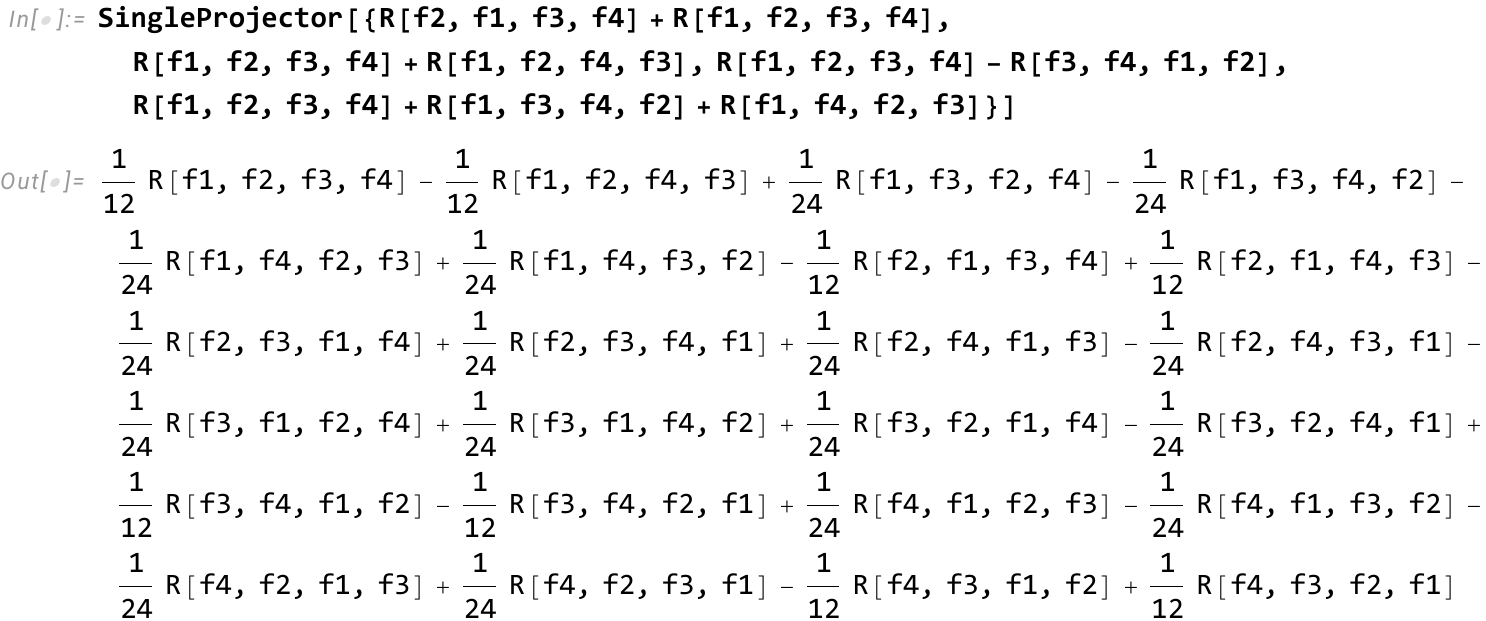}

\end{codeExample}

Note that Young symmetrizers are in general not hermitian (see for
example \cite{Keppeler:2013yla} and references contained therein):

\begin{codeExample}

\includegraphics[scale=0.62]{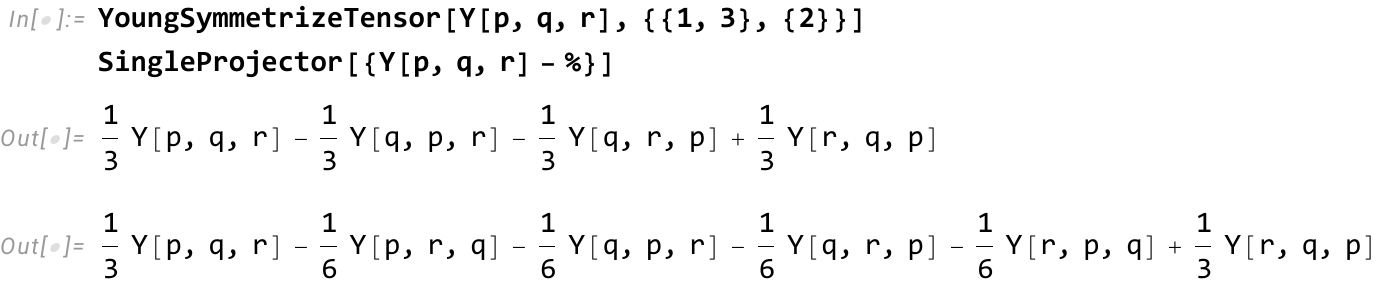}

\end{codeExample}

\subsection{Comparing sets of symmetry relations}

\begin{codeSyntax}
SameEquationsQ[<null equations 1>, <null equations 2>]
\tcblower
Compares the two sets of symmetry conditions (which may contain one or more tensor heads). The output is a string which identifies one out of 4 possible cases: (a) the equations are identical; (b) equations \#1 are more restrictive than equations \#2; (c) equations \#2 are more restrictive than equations \#1; (d) none of these apply (equations \#1 and \#2 are different).
\end{codeSyntax}

It might be important to know if two sets of equations, involving
tensors and their permutations, are the same or not. This can be checked
with the function \texttt{SameEquationsQ}. Note that the equations
can contain one of more tensors.

As a very simple example, the equation 
\begin{equation}
P_{k_{1}k_{2}}=Q_{k_{1}k_{2}}\label{eq:eq1}
\end{equation}
is equivalent to the following two equations:
\begin{equation}
P_{k_{1}k_{2}}+P_{k_{2}k_{1}}=Q_{k_{1}k_{2}}+Q_{k_{2}k_{1}}\textrm{ and }P_{k_{1}k_{2}}-P_{k_{2}k_{1}}=Q_{k_{1}k_{2}}-Q_{k_{2}k_{1}}\,.\label{eq:eq2}
\end{equation}
On the other hand, (\ref{eq:eq1}) is more restrictive than
\begin{equation}
P_{k_{1}k_{2}}+P_{k_{2}k_{1}}=Q_{k_{1}k_{2}}+Q_{k_{2}k_{1}}
\end{equation}
which only forces the symmetric part of the two tensors to be the
same. Finally
\begin{equation}
P_{k_{1}k_{2}}=Q_{k_{2}k_{1}}
\end{equation}
in neither equal, nor more restrictive, nor included in condition
(\ref{eq:eq1}), so in this sense one can say that is altogether different/unrelated
to (\ref{eq:eq1}). These comparisons can be performed with the following
code: 

\begin{codeExample}

\includegraphics[scale=0.62]{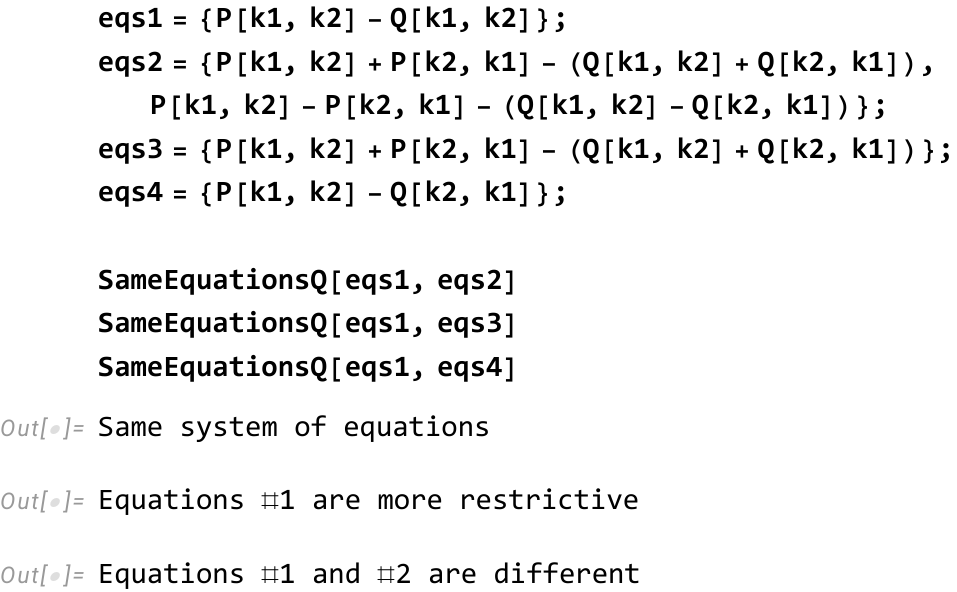}

\end{codeExample}

As a further example, one can check that $R_{abcd}=-R_{abdc}$ and
$-R_{bacd}+R_{acdb}+R_{adbc}=0$ are the same as the set of equations
in (\ref{eq:R-sym}) for the Riemann tensor:

\begin{codeExample}

\includegraphics[scale=0.62]{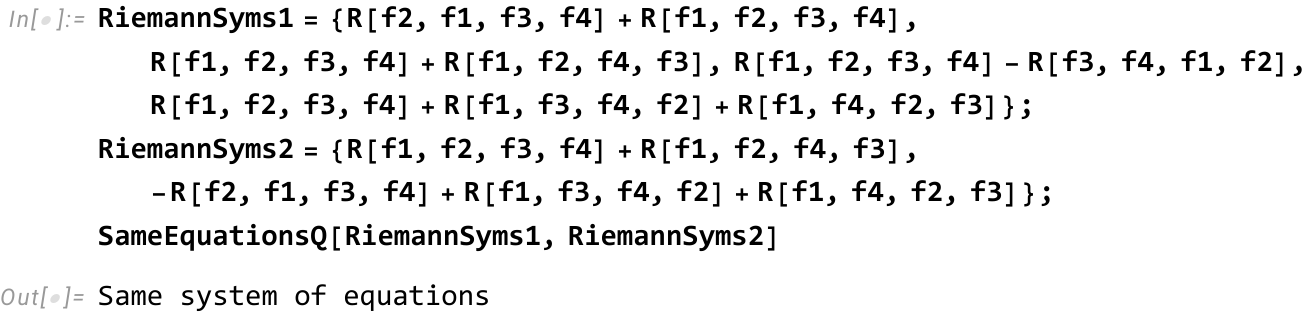}

\end{codeExample}

In fact, with \texttt{SingleProjector} one can express these symmetries
as a single equation:

\begin{codeExample}

\includegraphics[scale=0.62]{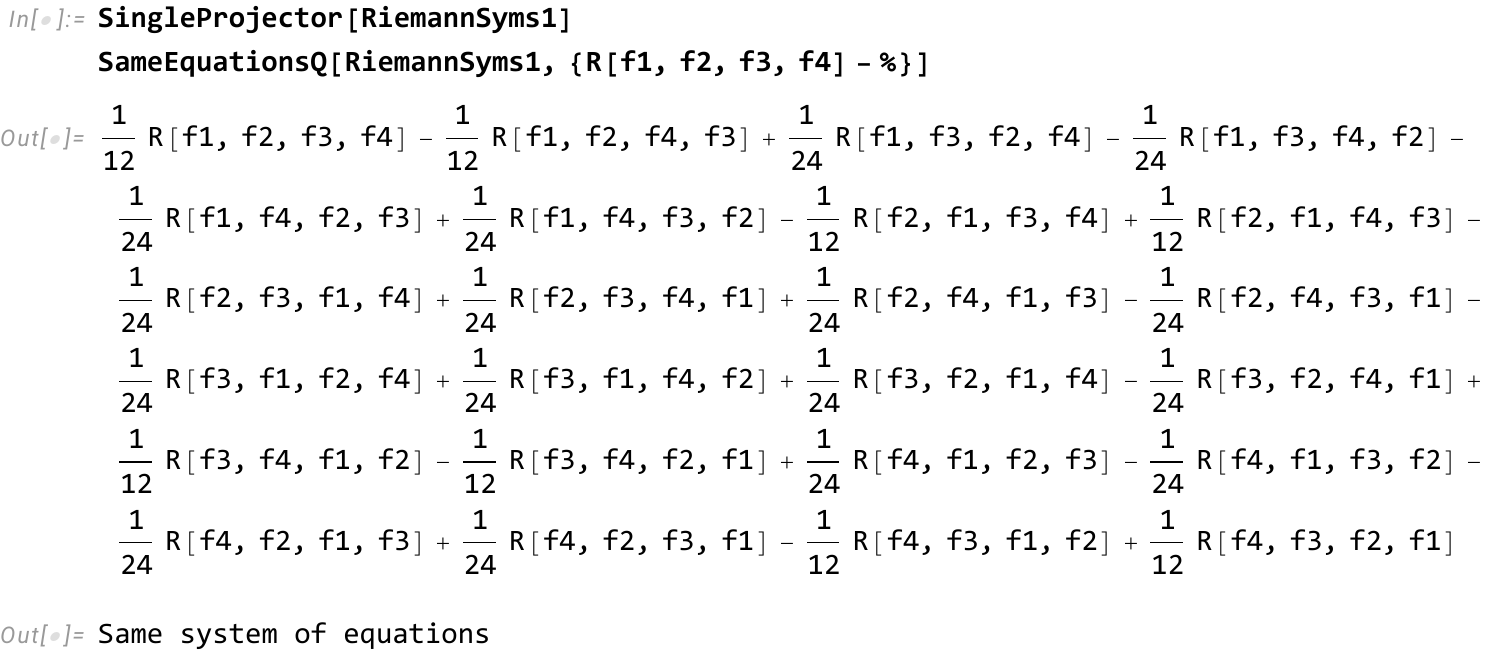}

\end{codeExample}

\subsection{\label{subsec:SymmetriesOfNumericalTensor}Symmetries of a numerical
tensor}

\begin{codeSyntax}
SymmetriesOfNumericalTensor[<numerical tensor>]
\tcblower
Returns a list of null equations encoding the symmetries of the given tensor, whose components are named "tensor"[id1,id2,...] in the output.
\end{codeSyntax}

One might want to know the symmetries of a known tensor, whose entries
are just numbers. The function \texttt{SymmetriesOfNumericalTensor}
returns a list of null expressions, which taken together contain all
the symmetry information of a given numerical tensor. Rather than
a single long expression, the result is often a list of several short
null conditions applicable to the tensor (if desired, these can be
condensed into a single one with the help to the \texttt{SingleProjector}
function). Here are some examples:

\begin{codeExample}

\includegraphics[scale=0.62]{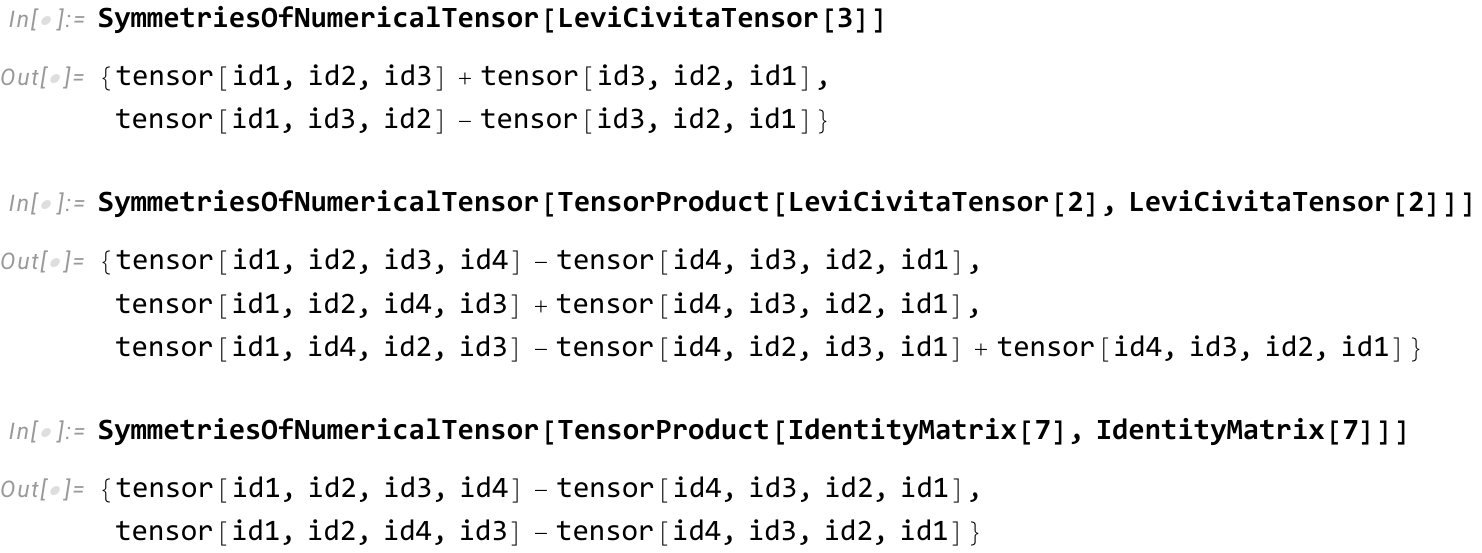}

\end{codeExample}

\section{\label{sec:Tensor-symmetries-beyond}Tensor symmetries beyond Young
symmetrizers}

There is a widespread understanding that the symmetry of a tensor
can always be described with Young symmetrizers. In this appendix
I will argue why this is not the case. The following discussion is
generalizable to any tensor, but for concreteness and simplicity let
us consider a 3-index tensor $T_{ijk}$, where each index takes $m$
values. We want to consider all symmetries that $T$ may possibly
have. Such symmetries can be expressed as 
\begin{equation}
P\left(T_{ijk}\right)=T_{ijk}\label{eq:P}
\end{equation}
or equivalently $\left(P-e\right)\left(T_{ijk}\right)=0$, with $P=P^{2}$
being some member of the 6-dimensional algebra of $S_{3}$:
\begin{equation}
P=x_{1}e+x_{2}\left(12\right)+x_{3}\left(13\right)+x_{4}\left(23\right)+x_{5}\left(123\right)+x_{6}\left(132\right)\,.\label{eq:S3-algebra}
\end{equation}
The symmetries of a tensor might be given as a list of several (rather
than just one) equations of this type.  However, one can always
translate those constraints into a single equation of the form (\ref{eq:P})
(and indeed this can be achieved with the function \texttt{SingleProjector}
presented in this work).\ytableausetup{boxsize=0.4em} 

Returning to the $T_{ijk}$ tensor, its $m^{3}$ components can then
be split in the following parts:
\begin{itemize}
\item A symmetric part $T_{a}^{\ydiagram{3}}$ with $\frac{1}{6}m\left(m+1\right)\left(m+2\right)$
components. It can be projected out with the Young symmetrizer $Y_{\ydiagram{3}}=\frac{1}{6}\left[e+\left(12\right)+\left(13\right)+\left(23\right)+\left(123\right)+\left(132\right)\right]$.
\item An antisymmetric part $T_{a}^{\ydiagram{1,1,1}}$ with $\frac{1}{6}m\left(m-1\right)\left(m-2\right)$
components. It can be projected out with the Young symmetrizer\ytableausetup{boxsize=0.4em} 
$Y_{\ydiagram{1,1,1}}=\frac{1}{6}\left[e-\left(12\right)-\left(13\right)-\left(23\right)+\left(123\right)+\left(132\right)\right]$.
\item A part with mixed symmetry $\ydiagram{2,1}$, having a total of $\frac{2}{3}\left(m+1\right)m\left(m-1\right)$
components. It can be projected out by the sum of Young symmetrizers
$Y_{\ydiagram{2,1}}^{(1)}+Y_{\ydiagram{2,1}}^{(2)}$ with $Y_{\ydiagram{2,1}}^{(1,2)}$
given below in equations (\ref{eq:YS-mix-1}) and (\ref{eq:YS-mix-2}).
\end{itemize}
Since $\ydiagram{2,1}$ is a 2-dimensional representation of the
permutation group, one can further split the space of mixed-symmetry
components in two, each with $\frac{1}{3}\left(m+1\right)m\left(m-1\right)$
components: $T_{i,a}^{\ydiagram{2,1}}$ with $i=1,2$ and $a=1,\cdots,\frac{1}{3}\left(m+1\right)m\left(m-1\right)$.
Combinations of the six permutations act on the first index ($i$)
only. In fact, one can make arbitrary transformations in this 2-dimensional
space spanned by the $i$ index, meaning that with a suitable combination
$P$ of the form (\ref{eq:S3-algebra}) applied to the tensor $T$
one can achieve any linear transformation
\begin{equation}
\left(\begin{array}{c}
T_{1,a}^{\ydiagram{2,1}}\\
T_{2,a}^{\ydiagram{2,1}}
\end{array}\right)\rightarrow\overset{M}{\overbrace{\left(\begin{array}{cc}
M_{11} & M_{12}\\
M_{21} & M_{22}
\end{array}\right)}}\cdot\left(\begin{array}{c}
T_{1,a}^{\ydiagram{2,1}}\\
T_{2,a}^{\ydiagram{2,1}}
\end{array}\right)\,.\label{eq:M}
\end{equation}
Note that each of the 4 $M_{ij}$'s is associated with an element
(\ref{eq:S3-algebra}) of the algebra of $S_{3}$; together with $Y_{\ydiagram{3}}$
and $Y_{\ydiagram{1,1,1}}$ they form a basis for this 6-dimensional
algebra.

There is an arbitrariness in defining a basis for the 2-dimension
space $\left\{ T_{1,a}^{\ydiagram{2,1}},T_{2,a}^{\ydiagram{2,1}}\right\} $;
a convenient way of fixing it is with Young symmetrizers, such as
\begin{align}
Y_{\ydiagram{2,1}}^{(1)} & =\frac{1}{3}\left[e+\left(12\right)\right]\left[e-\left(13\right)\right]=\frac{1}{3}\left[e+\left(12\right)-\left(13\right)-\left(132\right)\right]\,,\label{eq:YS-mix-1}\\
Y_{\ydiagram{2,1}}^{(2)} & =\frac{1}{3}\left[e+\left(13\right)\right]\left[e-\left(12\right)\right]=\frac{1}{3}\left[e+\left(13\right)-\left(12\right)-\left(123\right)\right]\,.\label{eq:YS-mix-2}
\end{align}
One can then define $T_{i,a}^{\ydiagram{2,1}}$ to be such that\footnote{Actually the two Young projectors only define $\left\{ T_{1,\alpha}^{\ydiagram{2,1}},T_{2,\alpha}^{\ydiagram{2,1}}\right\} $
up to (two) multiplicative factors.}
\begin{equation}
Y_{\ydiagram{2,1}}^{(1)}\left(\begin{array}{c}
T_{1,a}^{\ydiagram{2,1}}\\
T_{2,a}^{\ydiagram{2,1}}
\end{array}\right)=\left(\begin{array}{cc}
1 & 0\\
0 & 0
\end{array}\right)\cdot\left(\begin{array}{c}
T_{1,a}^{\ydiagram{2,1}}\\
T_{2,a}^{\ydiagram{2,1}}
\end{array}\right)\;\textrm{and}\;Y_{\ydiagram{2,1}}^{(2)}\left(\begin{array}{c}
T_{1,a}^{\ydiagram{2,1}}\\
T_{2,a}^{\ydiagram{2,1}}
\end{array}\right)=\left(\begin{array}{cc}
0 & 0\\
0 & 1
\end{array}\right)\cdot\left(\begin{array}{c}
T_{1,a}^{\ydiagram{2,1}}\\
T_{2,a}^{\ydiagram{2,1}}
\end{array}\right)\,.
\end{equation}
Therefore, a $Y_{\ydiagram{2,1}}^{(1)}$-symmetric tensor $T$, $T_{ijk}=Y_{\ydiagram{2,1}}^{(1)}\left(T_{ijk}\right)$,
will not have the $T_{2,a}^{\ydiagram{2,1}}$ part (nor the $T_{a}^{\ydiagram{3},\ydiagram{1,1,1}}$
parts). On the other hand, if $T_{ijk}=Y_{\ydiagram{2,1}}^{(2)}\left(T_{ijk}\right)$
the tensor $T$ will have $T_{1,a}^{\ydiagram{2,1}}$ zeroed out.
Clearly in both the cases the tensor is no longer general but rather
has a special, symmetric form.

The crucial point is that one can achieve arbitrary matrices $M$
in equation (\ref{eq:M}), including off-diagonal ones, so there are
symmetric tensors which cannot be described only with the projectors
$Y_{\ydiagram{2,1}}^{(1,2)}$. To make this observation more concrete,
first note that the transformation in (\ref{eq:M}) can be achieved
with the following element of the $S_{3}$ algebra:
\begin{align}
P_{M} & \equiv\frac{M_{11}+M_{22}}{3}e+\frac{M_{11}+M_{12}-M_{22}}{3}\left(12\right)+\frac{-M_{11}+M_{21}+M_{22}}{3}\left(13\right)+\frac{-M_{12}-M_{21}}{3}\left(23\right)\nonumber \\
 & +\frac{M_{12}-M_{21}-M_{22}}{3}\left(123\right)+\frac{-M_{11}-M_{12}+M_{21}}{3}\left(132\right)\,.
\end{align}
For any two matrices $M$ and $N$, one can check that, as it should,
$P_{M}P_{N}=P_{MN}$. Furthermore $Y_{\ydiagram{3}}P_{M}=Y_{\ydiagram{1,1,1}}P_{M}=0$.

For completeness, as a final step let us find all 2 by 2 matrices
$M$ which are projectors, as we want to apply $P_{M}$ to the $T$
tensor: $P_{M}\left(T_{ijk}\right)=T_{ijk}$. One trivial possibility
is $M=\boldsymbol{0}$ which zeroes out the full $\ydiagram{2,1}$
space of the $T$ tensor. The other trivial possibility is $M=\boldsymbol{1}$,
which keeps all this space; note that $P_{\boldsymbol{1}}$ can be
expressed using the Young projectors: $P_{\boldsymbol{1}}=Y_{\ydiagram{2,1}}^{(1)}+Y_{\ydiagram{2,1}}^{(2)}$. 

The more interesting scenario for the present discussion is when the
eigenvalues of $M$ are non-degenerate. We pick an eigenvector $\left(\cos\alpha,\sin\alpha\right)^{T}$
for the eigenvalue $1$ and another $\left(-\sin\beta,\cos\beta\right)^{T}$
for the eigenvalue $0$, yielding the matrix
\begin{equation}
M=\frac{1}{\cos(\alpha-\beta)}\left(\begin{array}{cc}
\cos\alpha\cos\beta & \cos\alpha\sin\beta\\
\sin\alpha\cos\beta & \sin\alpha\sin\beta
\end{array}\right)\,.\label{eq:M2}
\end{equation}
For this particular $M$, 
\begin{equation}
P_{M}\left(T_{ijk}\right)=T_{ijk}\Rightarrow\left(\begin{array}{c}
T_{1,a}^{\ydiagram{2,1}}\\
T_{2,a}^{\ydiagram{2,1}}
\end{array}\right)\propto\left(\begin{array}{c}
\cos\alpha\\
\sin\alpha
\end{array}\right)\textrm{ i.e. }T_{2,a}^{\ydiagram{2,1}}=\tan\alpha\,T_{1,a}^{\ydiagram{2,1}}\,.\label{eq:tan-alpha}
\end{equation}
Note that while the value of $P_{M}$ changes with $\beta$, this
angle does not impact the tensor $T$ itself. Also notice that $P_{M}$
does not really need to be a projector: $P_{M}\left(T_{ijk}\right)=T_{ijk}$
will kill any component of $T$ which is not associated to an eigenvalue
1, hence the exact value of the other eigenvalues $\neq1$ are irrelevant.
In fact, the only thing that matters are the eigenvectors of $P_{M}$
associated the eigenvalue 1.

But the most important observation is that one can have symmetric
tensors whose symmetry cannot be expressed with Young projectors.
The latter can be used to set $T_{1,a}^{\ydiagram{2,1}}=0$ or $T_{2,a}^{\ydiagram{2,1}}=0$,
but it is clear that one can have arbitrary relations between $T_{1,a}^{\ydiagram{2,1}}$
and $T_{2,a}^{\ydiagram{2,1}}$.

The discussion above holds true for tensors with higher rank, and
indeed the insufficiency of Young symmetrizers becomes more acute
as we consider larger irreducible representations of the permutation
group. For example, if $T$ has four indices, we can split it into
the 10 parts indicated below, and the elements of the 24-dimensional
algebra of $S_{4}$ act on it as follows: 

\begin{equation}
\left(\begin{array}{c}
T_{a}^{\ydiagram{4}}\\
T_{1,a}^{\ydiagram{3,1}}\\
T_{2,a}^{\ydiagram{3,1}}\\
T_{3,a}^{\ydiagram{3,1}}\\
T_{1,a}^{\ydiagram{2,2}}\\
T_{2,a}^{\ydiagram{2,2}}\\
T_{1,a}^{\ydiagram{2,1,1}}\\
T_{2,a}^{\ydiagram{2,1,1}}\\
T_{3,a}^{\ydiagram{2,1,1}}\\
T_{a}^{\ydiagram{1,1,1}}
\end{array}\right)\rightarrow\left(\begin{array}{cccccccccc}
\times\\
 & \times & \times & \times\\
 & \times & \times & \times\\
 & \times & \times & \times\\
 &  &  &  & \times & \times\\
 &  &  &  & \times & \times\\
 &  &  &  &  &  & \times & \times & \times\\
 &  &  &  &  &  & \times & \times & \times\\
 &  &  &  &  &  & \times & \times & \times\\
 &  &  &  &  &  &  &  &  & \times
\end{array}\right)\cdot\left(\begin{array}{c}
T_{a}^{\ydiagram{4}}\\
T_{1,a}^{\ydiagram{3,1}}\\
T_{2,a}^{\ydiagram{3,1}}\\
T_{3,a}^{\ydiagram{3,1}}\\
T_{1,a}^{\ydiagram{2,2}}\\
T_{2,a}^{\ydiagram{2,2}}\\
T_{1,a}^{\ydiagram{2,1,1}}\\
T_{2,a}^{\ydiagram{2,1,1}}\\
T_{3,a}^{\ydiagram{2,1,1}}\\
T_{a}^{\ydiagram{1,1,1}}
\end{array}\right)\,,
\end{equation}
where each cross can take any value (note that there are precisely
$24=4!$ of them). If, for example, we look at the 3-dimensional $\ydiagram{3,1}$
subspace, we find that the three Young symmetrizers are represented
by the matrices
\begin{equation}
\left(\begin{array}{ccc}
1 & 0 & 0\\
0 & 0 & 0\\
0 & 0 & 0
\end{array}\right),\left(\begin{array}{ccc}
0 & 0 & 0\\
0 & 1 & 0\\
0 & 0 & 0
\end{array}\right),\left(\begin{array}{ccc}
0 & 0 & 0\\
0 & 0 & 0\\
0 & 0 & 1
\end{array}\right)
\end{equation}
and again they are insufficient to express every conceivable tensor
symmetry. Take for instance the symmetry $T_{1,a}^{\ydiagram{3,1}}=T_{2,a}^{\ydiagram{3,1}}+2T_{3,a}^{\ydiagram{3,1}}$:
none of the $T_{i,a}^{\ydiagram{3,1}}$ is null so this kind of relation
cannot be obtained with Young symmetrizers, and yet it can be achieved
with the matrix
\begin{equation}
M=\left(\begin{array}{ccc}
1 & 0 & 0\\
0 & 1 & 0\\
\frac{1}{2} & -\frac{1}{2} & 0
\end{array}\right)
\end{equation}
since it has eigenvectors $\left(1,1,0\right)^{T}$ and $\left(2,0,1\right)^{T}$
associated to the eigenvalue 1.


\begin{thebibliography}{10}
	\providecommand{\url}[1]{\texttt{#1}}
	\providecommand{\urlprefix}{URL }
	\providecommand{\eprint}[2][]{\url{#2}}
	
	\bibitem{Einstein:1916vd}
	A.~Einstein, \emph{{The foundation of the general theory of relativity.}},
	\MYhref[journalLinks]{http://dx.doi.org/10.1002/andp.19163540702}{Annalen
		Phys.
	}\MYhref[journalLinks]{http://dx.doi.org/10.1002/andp.19163540702}{\textbf{49}
		(1916) 7 769--822}.
	
	\bibitem{Peeters:2018dyg}
	K.~Peeters, \emph{{Cadabra2: computer algebra for field theory revisited}},
	\MYhref[journalLinks]{http://dx.doi.org/10.21105/joss.01118}{J. Open Source
		Softw.
	}\MYhref[journalLinks]{http://dx.doi.org/10.21105/joss.01118}{\textbf{3}
		(2018) 32 1118}.
	
	\bibitem{Weinberg:1979sa}
	S.~Weinberg, \emph{{Baryon and lepton nonconserving processes}},
	\MYhref[journalLinks]{http://dx.doi.org/10.1103/PhysRevLett.43.1566}{Phys.
		Rev. Lett.
	}\MYhref[journalLinks]{http://dx.doi.org/10.1103/PhysRevLett.43.1566}{\textbf{43}
		(1979) 1566--1570}.
	
	\bibitem{Wilczek:1979hc}
	F.~Wilczek and A.~Zee, \emph{{Operator Analysis of Nucleon Decay}},
	\MYhref[journalLinks]{http://dx.doi.org/10.1103/PhysRevLett.43.1571}{Phys.
		Rev. Lett.
	}\MYhref[journalLinks]{http://dx.doi.org/10.1103/PhysRevLett.43.1571}{\textbf{43}
		(1979) 1571--1573}.
	
	\bibitem{Abbott:1980zj}
	L.~F. Abbott and M.~B. Wise, \emph{{The effective hamiltonian for nucleon
			decay}},
	\MYhref[journalLinks]{http://dx.doi.org/10.1103/PhysRevD.22.2208}{Phys. Rev.
	}\MYhref[journalLinks]{http://dx.doi.org/10.1103/PhysRevD.22.2208}{\textbf{D22}
		(1980) 2208}.
	
	\bibitem{Babu:2001ex}
	K.~S. Babu and C.~N. Leung, \emph{{Classification of effective neutrino mass
			operators}},
	\MYhref[journalLinks]{http://dx.doi.org/10.1016/S0550-3213(01)00504-1}{Nucl.
		Phys. B
	}\MYhref[journalLinks]{http://dx.doi.org/10.1016/S0550-3213(01)00504-1}{\textbf{619}
		(2001) 667--689},
	\MYhref[eprintLinks]{http://arxiv.org/abs/hep-ph/0106054}{{\ttfamily
			arXiv:hep-ph/0106054}}.
	
	\bibitem{deGouvea:2007qla}
	A.~de~Gouvea and J.~Jenkins, \emph{{A survey of lepton number violation via
			effective operators}},
	\MYhref[journalLinks]{http://dx.doi.org/10.1103/PhysRevD.77.013008}{Phys.
		Rev. D
	}\MYhref[journalLinks]{http://dx.doi.org/10.1103/PhysRevD.77.013008}{\textbf{77}
		(2008) 013008},
	\MYhref[eprintLinks]{http://arxiv.org/abs/0708.1344}{{\ttfamily
			arXiv:0708.1344 [hep-ph]}}.
	
	\bibitem{Fonseca:2018aav}
	R.~M. Fonseca and M.~Hirsch, \emph{{$\Delta L \ge 4$ lepton number violating
			processes}},
	\MYhref[journalLinks]{http://dx.doi.org/10.1103/PhysRevD.98.015035}{Phys.
		Rev. D
	}\MYhref[journalLinks]{http://dx.doi.org/10.1103/PhysRevD.98.015035}{\textbf{98}
		(2018) 1 015035},
	\MYhref[eprintLinks]{http://arxiv.org/abs/1804.10545}{{\ttfamily
			arXiv:1804.10545 [hep-ph]}}.
	
	\bibitem{Gargalionis:2020xvt}
	J.~Gargalionis and R.~R. Volkas, \emph{{Exploding operators for Majorana
			neutrino masses and beyond}},
	\MYhref[journalLinks]{http://dx.doi.org/10.1007/JHEP01(2021)074}{JHEP
	}\MYhref[journalLinks]{http://dx.doi.org/10.1007/JHEP01(2021)074}{\textbf{01}
		(2021) 074}, \MYhref[eprintLinks]{http://arxiv.org/abs/2009.13537}{{\ttfamily
			arXiv:2009.13537 [hep-ph]}}.
	
	\bibitem{Tung-book}
	W.~Tung, \emph{Group theory in Physics}, World Scientific (1985).
	
	\bibitem{Bolotin:2013qgr}
	D.~A. Bolotin and S.~V. Poslavsky, \emph{{Introduction to Redberry: a computer
			algebra system designed for tensor manipulation}}  (2013),
	\MYhref[eprintLinks]{http://arxiv.org/abs/1302.1219}{{\ttfamily
			arXiv:1302.1219 [cs.SC]}}.
	
	\bibitem{Ilyin:1996otf}
	V.~A. Ilyin and A.~P. Kryukov, \emph{{ATENSOR -- REDUCE program for tensor
			simplification}},
	\MYhref[journalLinks]{http://dx.doi.org/10.1016/0010-4655(96)00060-4}{Comput.
		Phys. Commun.
	}\MYhref[journalLinks]{http://dx.doi.org/10.1016/0010-4655(96)00060-4}{\textbf{96}
		(1996) 1 36--52},
	\MYhref[eprintLinks]{http://arxiv.org/abs/1811.05409}{{\ttfamily
			arXiv:1811.05409 [cs.SC]}}.
	
	\bibitem{Price:2022wlt}
	D.~Price, K.~Peeters and M.~Zamaklar, \emph{{Hiding canonicalisation in tensor
			computer algebra}}  (2022),
	\MYhref[eprintLinks]{http://arxiv.org/abs/2208.11946}{{\ttfamily
			arXiv:2208.11946 [cs.SC]}}.
	
	\bibitem{Martin-Garcia:2008ysv}
	J.~M. Mart\'\i{}n-Garc\'\i{}a, \emph{{xPerm: fast index canonicalization for
			tensor computer algebra}},
	\MYhref[journalLinks]{http://dx.doi.org/10.1016/j.cpc.2008.05.009}{Comput.
		Phys. Commun.
	}\MYhref[journalLinks]{http://dx.doi.org/10.1016/j.cpc.2008.05.009}{\textbf{179}
		(2008) 8 597--603},
	\MYhref[eprintLinks]{http://arxiv.org/abs/0803.0862}{{\ttfamily
			arXiv:0803.0862 [cs.SC]}}.
	
	\bibitem{xAct}
	J.~M. Mart\'\i{}n-Garc\'\i{}a et~al., \emph{{xAct: Efficient tensor computer
			algebra for the Wolfram Language}}, \urlprefix\url{http://www.xact.es/}.
	
	\bibitem{Fonseca:2020vke}
	R.~M. Fonseca, \emph{{GroupMath: A Mathematica package for group theory
			calculations}},
	\MYhref[journalLinks]{http://dx.doi.org/10.1016/j.cpc.2021.108085}{Comput.
		Phys. Commun.
	}\MYhref[journalLinks]{http://dx.doi.org/10.1016/j.cpc.2021.108085}{\textbf{267}
		(2021) 108085},
	\MYhref[eprintLinks]{http://arxiv.org/abs/2011.01764}{{\ttfamily
			arXiv:2011.01764 [hep-th]}}.
	
	\bibitem{Algorithms-for-computer-algebra}
	K.~O. Geddes, S.~R. Czapor and G.~Labahn, \emph{Algorithms for computer
		algebra}, Kluwer Academic Publishers, USA (1992).
	
	\bibitem{Fonseca:workInProgress}
	R.~Fonseca, P.~Olgoso and J.~Santiago, \emph{{Renormalisation of general
			effective field theories: formalism and renormalisation of bosonic operators
		} (in preparation)} .
	
	\bibitem{Fonseca:2019yya}
	R.~M. Fonseca, \emph{{Enumerating the operators of an effective field theory}},
	\MYhref[journalLinks]{http://dx.doi.org/10.1103/PhysRevD.101.035040}{Phys.
		Rev. D
	}\MYhref[journalLinks]{http://dx.doi.org/10.1103/PhysRevD.101.035040}{\textbf{101}
		(2020) 3 035040},
	\MYhref[eprintLinks]{http://arxiv.org/abs/1907.12584}{{\ttfamily
			arXiv:1907.12584 [hep-ph]}}.
	
	\bibitem{Keppeler:2013yla}
	S.~Keppeler and M.~Sj\"odahl, \emph{{Hermitian Young operators}},
	\MYhref[journalLinks]{http://dx.doi.org/10.1063/1.4865177}{J. Math. Phys.
	}\MYhref[journalLinks]{http://dx.doi.org/10.1063/1.4865177}{\textbf{55}
		(2014) 021702},
	\MYhref[eprintLinks]{http://arxiv.org/abs/1307.6147}{{\ttfamily
			arXiv:1307.6147 [math-ph]}}.
	
\end{thebibliography}
\end{document}